\documentclass{aa}  
\usepackage{graphicx}

\usepackage{txfonts}
\usepackage{tabularx}
\usepackage{amsfonts}
\usepackage{bbold}
\usepackage{color}
\usepackage{transparent}
\usepackage{hyperref}
\usepackage{transparent}
\usepackage{rotating}
\usepackage{caption}
\usepackage{subcaption}

\usepackage{graphicx}	
\usepackage{amsmath}	
\usepackage{amssymb}	
\usepackage{tablefootnote}
\usepackage[flushleft]{threeparttable}
\usepackage{dblfloatfix} 
\usepackage{siunitx}
\usepackage{array}

\usepackage{physics}

\usepackage{multirow}
\usepackage{diagbox}
\usepackage[normalem]{ulem}

\usepackage{natbib}
\bibpunct{(}{)}{;}{a}{}{,} 

\usepackage{empheq}

\newcommand*{\bh}{BH\@\xspace}
\newcommand*{\bhs}{BHs\@\xspace}
\newcommand*{\eg}{e.g.\@\xspace}
\newcommand*{\ie}{i.e.\@\xspace}
\newcommand*{\aka}{a.k.a. \@\xspace}

\newcommand*{\sgr}{\mbox{Sgr A$^*$}\@\xspace}

\newcommand*{\smbh}{SMBH\@\xspace}

\newcommand*{\isco}{\text{ISCO}\@\xspace}

\newcommand*{\fidos}{FIDOs\@\xspace}

\newcommand*{\pp}{Paper \textrm{I}\@\xspace}

\newcommand*{\ileyk}{\textcolor[rgb]{0,0,0}} 

\usepackage[dvipsnames]{xcolor}

\makeatletter
\newcommand{\thickhline}{%
    \noalign {\ifnum 0=`}\fi \hrule height 1pt
    \futurelet \reserved@a \@xhline
}
\makeatother

\begin{document} 




     \title{Reconnection-driven flares in 3D black hole magnetospheres}
    \subtitle{A scenario for hot spots around Sagittarius A*}
    
   \author{I. El Mellah
          \inst{1,2,3}
          \and
          B. Cerutti
          \inst{1} 
          \and
          B. Crinquand
          \inst{4} 
          }

   \institute{Univ. Grenoble Alpes, CNRS, IPAG, 38000 Grenoble, France\\
              \email{ileyk.el-mellah@usach.cl}
         \and
         	Departamento de F\'{i}sica, Universidad de Santiago de Chile, Av. Victor Jara 3659, Santiago, Chile
          \and 
                Center for Interdisciplinary Research in Astrophysics and Space Exploration (CIRAS), USACH, Chile 
          \and 
                Department of Astrophysical Sciences, Peyton Hall, Princeton University, Princeton, New Jersey 08544, USA
    }
   \date{Received ...; accepted ...}

 
  \abstract
   {Low-luminosity supermassive and stellar-mass black holes (\bhs) may be embedded in a collisionless and highly magnetized plasma. They show non-thermal flares indicative of particles being accelerated up to relativistic speed by dissipative processes in the vicinity of the \bh. During near-infrared flares from the supermassive \bh Sagittarius A$^*$ (\sgr), the GRAVITY collaboration detected circular motion and polarization evolution which suggest the presence of transient synchrotron-emitting hot spots moving around the \bh.}
   {We study 3D reconnecting current layers in the magnetosphere of spinning \bhs to determine whether plasma-loaded flux ropes formed near the event horizon could reproduce the hot spot observations and help constraining the \bh spin.} 
   {We perform global 3D particle-in-cell simulations in Kerr spacetime of a pair plasma embedded in a strong and large-scale magnetic field originating in a perfectly conducting disk in prograde Keplerian rotation.}
   {A cone-shaped current layer develops which surrounds the twisted open magnetic field lines threading the event horizon. Spinning magnetic field lines coupling the disk to the \bh inflate and reconnect a few gravitational radii above the disk. This quasi-periodic cycle accelerates particles which accumulate in a few macroscopic flux ropes \ileyk{rotating with the outermost coupling magnetic field line. Once flux ropes detach, they propagate in the current layer following what appears as a rapidly opening spiral when seen face-on.} A single flux rope carries enough relativistic electrons and positrons to emit synchrotron radiation at levels suitable to reproduce the peak-luminosity of the flares of \sgr but it quickly fades away as it flows away.}
   {Our kinematic analysis \ileyk{of the flux ropes motion} favors a \bh spin of $0.65$ to $0.8$ for \sgr. The duration of the flares of \sgr can only be explained provided the underlying magnetic loop seeded in the disk mid-plane has a finite lifetime and azimuthal extension. In this scenario, the hot spot corresponds to a spinning arc along which multiple reconnection sites power the net emission as flux ropes episodically detach.}
   

   \maketitle


\section{Introduction}




Black holes (\bhs) are sources of transient non-thermal emission whose origin remains elusive. Being the closest and largest \bh as seen from Earth, Sagittarius A$^*$ (\sgr) is a privileged target to study the plasma dynamics in its immediate vicinity. Stringent constraints have been set on its mass from the orbits of stars in the nuclear cluster of the Milky Way \citep{Schoedel2009,Gillessen2009,Abuter2018c} and from the size of the \bh shadow imaged by \cite{Collaboration2022}. Direct imaging, astrometry and polarimetry of the emission from the surrounding plasma have progressively suggested that the \bh spin axis is seen at low inclination that is to say almost face-on \citep{Jimenez-Rosales2020,GravityCollaboration2021,Collaboration2022,Collaboration2022,Wielgus2022}. \cite{Collaboration2022} carried an extensive comparison of the observations of \sgr to general relativistic magneto-hydrodynamics (GRMHD) simulations. The only models which match all the observational constrains in the quiescent phase are those with positive \bh spins (0.5 to 0.94) magnetically-arrested disks \citep[MAD, in agreement with the larger scales simulations of][]{Ressler2020a} and low viewing angles ($<30^{\circ}$). Reasonable hopes can be nurtured that GRAVITY$+$ will be able to bring independent constraints from the Lens-Thirring precession of the orbit of stars even closer from the \bh than S2 \citep{Abuter2020,Abuter2021}.


On top of its quiescent emission, \sgr shows hour-long flares at near-infrared (NIR) and sometimes X-ray wavelengths on a daily basis \citep{Baganoff2001,Genzel2003,Ghez2004}. The similar time scales and recurrent simultaneity of NIR and X-ray flares suggest that they might be the outcome of a common mechanism \citep{Boyce2019}. A long-suspected culprit is the acceleration of electrons and positrons in magnetically reconnecting current sheets \citep{Yuan2004,Ball2018,Comisso2021}. When \sgr is in its quiescent state, the linearly polarized emission is well reproduced by a synchrotron emission from a population of relativistic electrons embedded in a strong and structured magnetic field near the \bh event horizon \citep{Bower2018}. During flares, a high degree of linear polarization is also reported \citep{Jimenez-Rosales2020}. The highly magnetized collisionless plasma which surrounds the \bh is \ileyk{likely} a fruitful environment for reconnection sites where electromagnetic energy is dissipated at maximal rates. This process typically yields non-thermal particle energy distribution \citep{Sironi2014,Werner2015,Werner2017,Rowan2017}, especially when synchrotron cooling is inefficient like around \sgr.


During three NIR flares, the ESO VLTI/GRAVITY instrument found that the centroid \ileyk{of the emission} moved clockwise around the \bh \cite{Abuter2018b}. This major discovery sheds light on the origin of these flares since it grants us simultaneous access to the kinematics and radiative power of the emitting region coined as the hot spot. The flares are brighter than the quiescent emission, so that we can safely assume that the centroid traces the motion of the emitting region \citep{GravityCollaboration2021}. \ileyk{In each case, the hot spot described an incomplete circle of radius 35 to 50$\mu$as (\ie 7 to 10 gravitational radii) in 15 to 30 minutes, and whose center was shifted by almost 50$\mu$as with respect to the projected position of \sgr.} Taken together, these kinematics constraints indicate that the motion may be super-Keplerian by $\sim$20-30\%. Further modeling of the hot spot motion by \cite{Baubock2020} found that the astrometric data was consistent with a viewing angle lower than $<40^{\circ}$ and a steady out-of-plane motion at 0.15$c$, with $c$ the speed of light. Although marginal, the latter result suggests that the hot spot might be launched from the basis of a jet with an upwards velocity component rather than purely orbiting in the equatorial plane of the \bh.


High resolution 3D ideal GRMHD simulations of accretion onto \bhs in the MAD state show recurrent relaxation of the magnetic flux piled up onto the event horizon \citep[see \eg][]{Tchekhovskoy2011,Dexter2020,Ripperda2022,Galishnikova2022}. This cycle is mediated by the episodic ejection of magnetic flux through highly magnetized bubbles whose envelope is prone to contain magnetically reconnecting current sheets in the disk mid-plane. The particles accelerated through this process reach high Lorentz factors and emit non-thermal synchrotron radiation which could be responsible for the flares from \sgr \citep{Scepi2022}. More generally, there are multiple studies which computed synthetic observables from a wide range of GRMHD simulations \citep{Dexter2020a,Cruz-Osorio2022,Vos2022,Wong2022,Collaboration2022,Chael2023} but they all rely on assumptions for the electron distribution function and the electron heating process since GRMHD is inherently unable to capture the micro-physics. In parallel, several numerical studies characterized the kinetic properties of the collisionless plasma in local simulations of turbulence and/or magnetic reconnection \citep{Meringolo2023,Zhdankin2023}. \ileyk{Current sheets are also thought to form out of the disk plane. There are empirical models of non-thermal flaring emission from particles accelerated through reconnection of magnetic loops rising above the disk due to turbulence and/or Keplerian shearing \citep{Yuan2009,Lin2023}. Finally, in GRMHD simulations of accretion of magnetic loops of opposite polarity, reconnection-induced flux ropes form in the envelope of the \bh jet \citep{Nathanail2020,Chashkina2021a,Nathanail2022}.}


In \cite{ElMellah2022}, hereafter \pp, we investigated a scenario where the main dissipation sites are not located in the disk plane but rather in the oblique layer separating the open magnetic field lines threading the event horizon and those anchored on the disk. To do so, we studied the properties of a \bh magnetosphere loaded with pair-plasma and where magnetic field lines couple the \bh to the disk. We performed 2D axisymmetric (hereafter, 2.5D) global particle-in-cell (PIC) simulations in Kerr metric with the \texttt{Zeltron} code \citep{Cerutti2013,Parfrey2019} for \bh spins from 0.6 to 0.99. Contrary to GRMHD, the PIC framework does not rely on a fluid or collisional approximation and can accurately describe dissipative processes in collisionless plasmas like magnetic reconnection. In agreement with previous conjectures, the shearing of the magnetic field lines induced by frame dragging in the \bh ergosphere leads to the formation of a Y-shaped magnetic field topology, or Y-ring, above the disk \cite{Uzdensky2005,DeGouveiaDalPino2005,Yuan2019b}. Beyond the Y-ring, a cone-shaped current sheet separates the inner jet from the disk magneto-centrifugal outflow. In this current sheet, vivid magnetic reconnection produces \ileyk{non-thermal particles} and macroscopic plasmoids form as the outermost closed magnetic field line episodically inflate and reconnect. 


By construction, the 2.5D numerical setup we designed in \pp was unable to capture non-axisymmetric features, \ileyk{a fortiori a localized hot spot moving around the black hole}. Furthermore, the polarity of the toroidal magnetic field component also reversed across the current sheet but we could only capture magnetic reconnection in the poloidal plane. More generally, magnetic reconnection is known to proceed in a quantitatively and qualitatively different manner in 3D and in 2D \citep{Werner2017,Comisso2019,Zhang2021,Werner2021}. In this respect, effects such as enhanced turbulence, the triggering of azimuthal instability modes and the shear of angular speed across the current sheet remained to be investigated in our model. 


In this paper, we relax the axisymmetric assumption and perform global 3D general relativistic PIC (GRPIC) simulations of a \bh magnetosphere in order to overcome the aforementioned limitations of our previous 2.5D setup. Hereafter and following the common denomination, we will refer to 2D plasma-loaded magnetic islands as plasmoids and to 3D plasma-loaded entwined magnetic field lines as flux ropes. Similarly, the 2D concepts of current sheets and X/Y-points are replaced in 3D by current layers and null-lines respectively. We pay special attention to the possibility that \ileyk{moving flux ropes} might manifest as synchrotron-bright hot spots. For the sake of computational affordability, we work with a dimensionless \bh spin of 0.99, but we extrapolate our results to lower spin values in the light of our past 2.5D simulations.


In Section\,\ref{sec:model}, we recall the main ingredients of our model and the conclusions we drew from our 2.5D simulations. Then, we present the 3D results we obtained in terms of magnetic topology, plasma properties and reconnection rate in Section\,\ref{sec:results}. In Section\,\ref{sec:discussion}, we perform an in-depth analysis of the kinematic and radiative properties of the hot spots that we identified in order to compare them to the observations and set constraints on \sgr's spin and on the physical mechanism responsible for the flares. We summarize our results and suggest follow-up work in Section\,\ref{sec:summary}.










\section{Model}
\label{sec:model}

\subsection{Physics and code}
\label{sec:code}

We study the three-dimensional dynamics of the electron/positron pair plasma and of the electromagnetic fields in the magnetosphere of a spinning \bh. The background Kerr metric is stationary and axisymmetric. It is determined (i) by the \bh mass, $M$, which sets the length scale via the gravitational radius $r_g=GM/c^2$, with $G$ the gravitational constant, and (ii) by the \bh dimensionless spin $a$ \citep{Kerr1963}. We rely on the $3+1$ formalism \citep{MacDonald1982,Komissarov2004a} and on the spherical Kerr-Schild coordinates $(t,r,\theta,\phi)$.

We use the GRPIC code \texttt{Zeltron} to advance in time the position and velocity of the particles along with the electric and magnetic fields, respectively $\mathbf{E}$ and $\mathbf{B}$, measured by the fiducial observers (\fidos, whose worldlines are orthogonal to spatial hypersurfaces of constant time coordinate) \citep{Cerutti2013,Parfrey2019,Crinquand2021}. We do not account for radiative drag forces in the particles' equation of motion, and neglect radiative cooling, a safe assumption for \sgr.


\subsection{Numerical setup}
\label{sec:setup}

The 3D grid on which electromagnetic fields are advanced and charges and currents are deposited has a resolution of 512$(r)\times$256$(\theta)\times$256$(\phi)$. It extends over $r\in[0.9r_h,30r_g]$, $\theta\in[\pi/96,\pi/2]$ and $\phi\in[0,\pi/2]$, with a logarithmic stretching in $r$, and $r_h$ is the radius of the event horizon. \ileyk{Our choice of $\phi$-extent stems from the need to capture the azimuthal dynamics (see Section\,\ref{sec:rec_rate}).} The \bh equatorial plane $\theta=\pi/2$ is assumed to be a plane of symmetry. 

\begin{figure}[!h]
\centering
\begin{subfigure}[b]{0.95\columnwidth}
   \includegraphics[width=0.99\columnwidth]{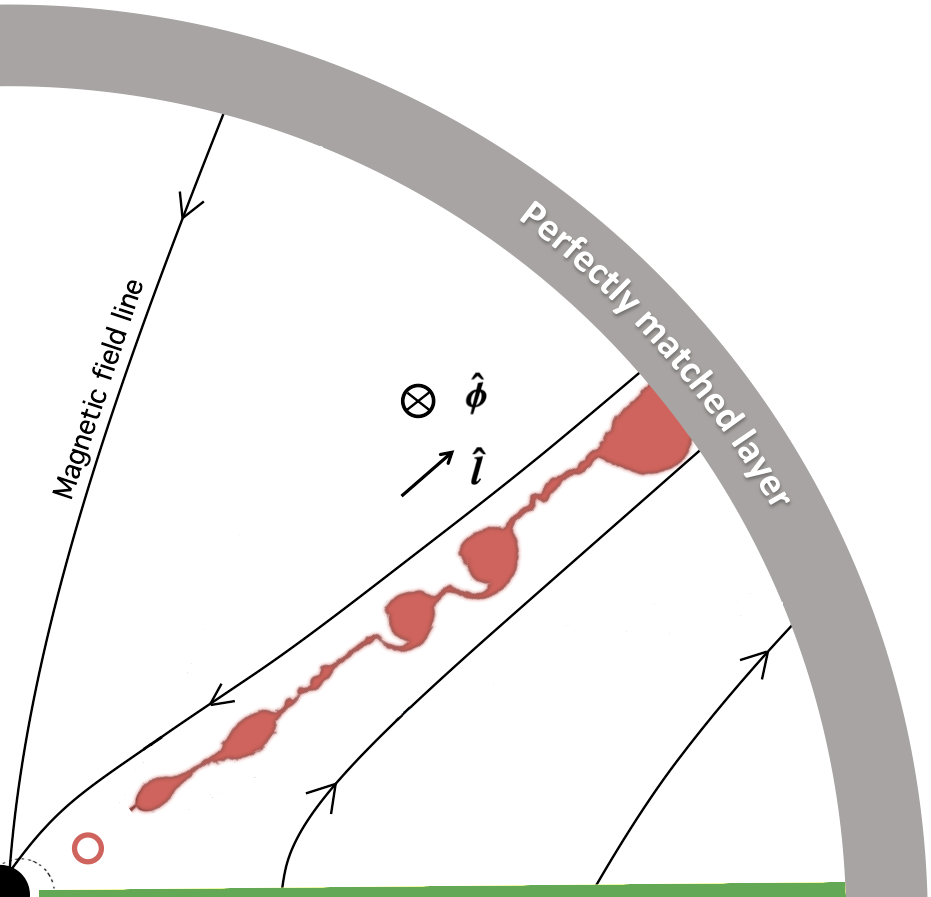}
\end{subfigure}
\begin{subfigure}[b]{0.95\columnwidth}
   \includegraphics[width=0.99\columnwidth]{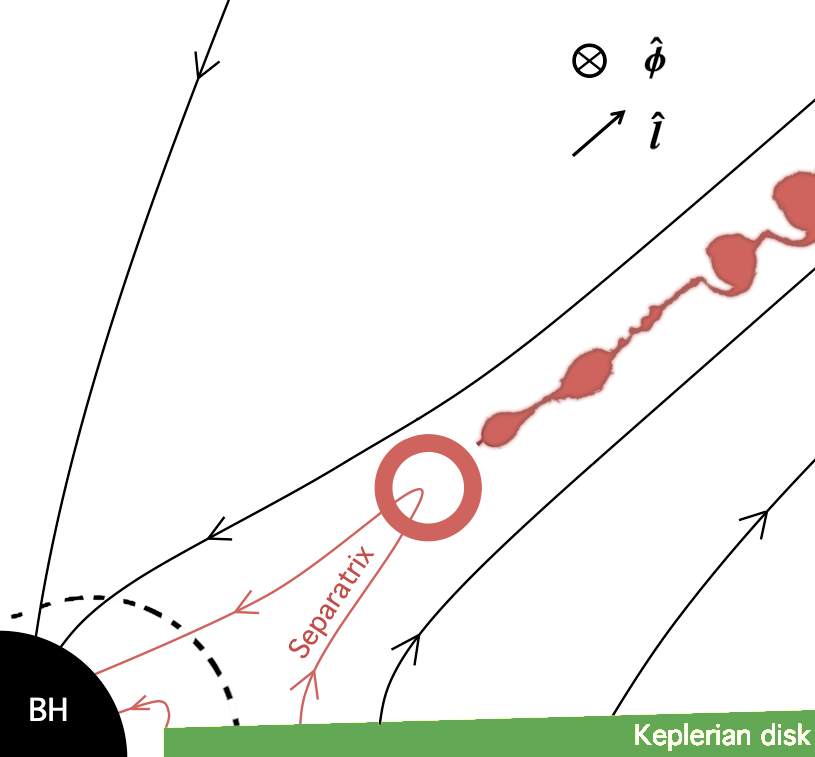}
\end{subfigure}
\caption{Sketches of the axisymmetric 2.5D \bh magnetosphere for a spin $a=0.99$. \textit{(top)} The black disk (resp. black dashed line) in the bottom left corner is the event horizon of size $r_h\sim 1.14r_g$ (resp. the ergosphere). The black lines represent open magnetic field lines. The absorbing layer for outer boundary conditions is shown in grey. The red circle locates the Y-point at the basis of the reconnecting current sheet in which plasmoids are visible in red. The green region is the Keplerian disk where magnetic field lines are anchored. \textit{(bottom)} Zoom in on the innermost region, with closed magnetic field lines coupling the disk to the \bh represented in red, along with the outermost closed magnetic field line (the separatrix).}
\label{fig:25D}
\end{figure}

\ileyk{We enforce zero-gradient boundary conditions at the inner edge, within the event horizon,} and we use an absorbing layer at the outer edge, from 27$r_g$ to 30$r_g$ \citep[\ileyk{see top panel in Figure\,\ref{fig:25D}} and][]{Cerutti2015}. At $\theta=\pi/96$, we apply reflective boundary conditions. Boundary conditions in $\phi$ are periodic and hereafter, figures where $\phi$ extends beyond the range [$0,\pi/2$] are obtained through duplication and concatenation. Between $\theta=\pi/2-\arctan\left(\epsilon\right)$ and $\theta=\pi/2$, we introduce a perfectly conducting disk of aspect ratio $\epsilon=5\%$. The disk is aligned with the \bh spin axis and in prograde rotation at the Keplerian angular speed
\begin{equation}
    \Omega_K(R)=\sqrt{\frac{GM}{r_g^3}}\frac{1}{\left(R/r_g\right)^{3/2}+a},
\end{equation}
with $R=r\sin(\theta)$ the projected distance to the spin axis. Within the innermost stable circular orbit (\isco), we set $\Omega_K=0$. \ileyk{In the disk, the magnetic field $\boldsymbol{B}$ is axisymmetric and the magnetic field lines are frozen and enforced to rotate at the angular speed of their footpoint in the disk mid-plane. The electric field in the local co-rotating frame cancels out since the disk is a perfect conductor, which sets the boundary conditions for the electric field.}

Observations suggest that the hot and dilute plasma in the immediate vicinity of weakly accreting \bhs such as \sgr might be highly magnetized. Its global dynamics are thus essentially force-free, with particles flowing along magnetic field lines and screening any parallel electric field which would be susceptible to accelerate them. To reproduce those conditions, we continuously inject pairs of particles at rest in the local FIDO frame throughout the simulation domain such as the plasma density $n$ never goes below three times \ileyk{the absolute local Goldreich-Julian density $|n_{\textrm{GJ}}|$ defined as \citep{Goldreich1969}
\begin{equation}
    \label{eq:GJ}
    |n_{\textrm{GJ}}|=\frac{|\boldsymbol{\Omega}\cdot\boldsymbol{B}|}{2\pi c},
\end{equation}
where $\boldsymbol{\Omega}$ is the local angular speed along the azimuthal direction.} It ensures that there are enough charges for the force-free regime to be achieved wherever dissipation does not take place. Another necessary condition of the force-free regime is
\begin{equation}
\label{eq:sigma}
    \sigma=\frac{|\mathbf{B}|^2}{4\pi n \Gamma m_e c^2}\gg 1,
\end{equation}
with $\sigma$ the magnetization parameter, \ileyk{$\Gamma$ the bulk Lorentz factor} and $m_e$ the mass of the electron. We achieve this regime while still resolving the plasma skin depth $\delta$ by working with a Larmor radius $R_L=10^{-5}r_g$. Although the corresponding magnetic field is 3 to 5 orders of magnitude lower than what is measured in \sgr, the hierarchy between scales is preserved. Furthermore, the underestimations of $|\mathbf{B}|$ and $n$ compensate in Eq.~\eqref{eq:sigma} such that the magnetization $\sigma\sim1,000$ \ileyk{is high enough to ensure that we work in the quasi-force-free regime $\sigma \gg 1$.}

Finally, we choose to work with a very high \bh spin of $a=0.99$ for computational convenience. Indeed, we showed in \pp that higher spin values led to a region of interest developing closer from the \bh and evolving on shorter time scales. Given the high cost of 3D global GRPIC simulations, it is a major advantage that we build upon to compare our 3D results to our previous 2.5D simulations and extrapolate them to lower spin values. With such a high spin, the inner edge of the disk lies at $\sim1.4r_g$, within the ergosphere and very close from the event horizon.

\begin{figure*}[t]
\centering
\includegraphics[width=1.9\columnwidth,height=6.5cm]{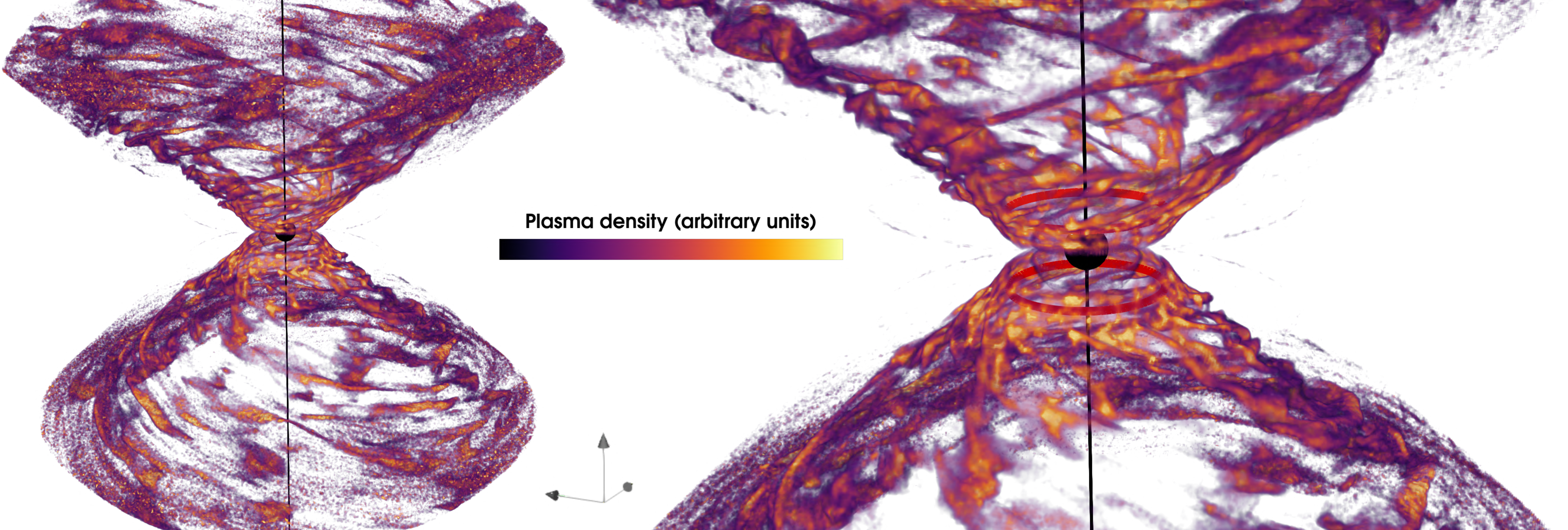}
\caption{3D volume rendering of plasma density in the current layer around the \bh event horizon (central black sphere). Flux ropes are stretched in the azimuthal direction and break up into visible overdense plasma filaments flowing outwards along the cone-shaped current layer. The black vertical line is the spin axis of the black hole while the two other coordinate axis in grey are in the equatorial plane. In the right panel (zoomed in), the red rings above and under the equatorial plane locate the Y-ring.}
\label{fig:plasma_density}
\end{figure*}

\subsection{Relaxation to 3D equilibrium}
\label{sec:relax}

The initial conditions we start from are the same as in \pp: a dipolar magnetic field in vacuum such as initially, all magnetic field lines thread the event horizon and are anchored in the steady Keplerian disk at their other end. \ileyk{The poloidal magnetic field profile in the disk mid-plane is thus $\propto R^{-3}$.} In order to speed up the relaxation of the initial conditions, we follow a two-step procedure toward a fully three-dimensional relaxed simulation.

\ileyk{In order to speed up relaxation of the initial conditions, we first work with a 3D setup with only 4 cells along $\phi$, extending from $\phi=0$ to $\phi=0.025$, and with periodic boundary conditions in $\phi$.} It sets a stringent constraint on the azimuthal extension of the structures such as the simulation is essentially axisymmetric and computationally affordable due to the low number of cells along $\phi$.

After $\sim 150r_g/c$, the setup has relaxed. Relaxation of the initial conditions proceeds along the same lines as in 2.5D: the spinning-up torques onto the magnetic field lines induced by the frame dragging in the ergosphere lead to an increase of the toroidal component of the magnetic field. Magnetic field loops whose outer footpoint is close enough from the \isco can catch up for this additional magnetic tension but beyond a critical distance, magnetic field lines necessarily open up. The higher the \bh spin, the closer the outermost closed magnetic field line (hereafter, the separatrix). 


This \bh rotation powered mechanism structures the magnetosphere in regions of distinct topology. \ileyk{In Figure\,\ref{fig:25D}, we show an illustration of the axisymmetric 2.5D configuration where the main features are represented.} In the innermost region between the \isco and the footpoint of the separatrix, closed magnetic field lines couple the \bh to the disk \ileyk{(bottom panel in Figure\,\ref{fig:25D})}. They carry energy and angular momentum between the two components \citep{Uzdensky2005,Yuan2019b,ElMellah2022}. Toward the pole, open and strongly twisted magnetic field lines thread the \bh and form the backbone of an electromagnetic jet \citep{Blandford1977}. Beyond the separatrix, the disk is threaded with inclined open magnetic field lines. From the Y-point at the intersection of these 3 regions, a current sheet forms between the jet and the disk outflow \ileyk{(red plasmoids in Figure\,\ref{fig:25D})}. In this layer, the force-free approximation breaks down and the magnetic field reconnects. 

Near the separatrix, a quasi-periodic cycle sets up: field lines progressively stretch out as the toroidal component increases, until the pinching of their outermost section forms a current sheet whose reconnection is triggered by the tearing instability. Through this process, magnetic islands episodically flow away from the Y-point in the current sheet. They correspond to higher plasma density regions commonly called plasmoids in 2D simulations of magnetic reconnection (visible in Fig\,\ref{fig:25D}) and flux ropes in 3D. In-between plasmoids, the current sheet reaches its minimal thickness, the skin depth, at X-points where particles are accelerated by the non-ideal electric field induced by reconnection. In 3D, the Y-point becomes a Y-ring above the disk.

Once this setup has relaxed, we unfold it around the spin axis over the full $\phi$ range by repeating the fields and particles information. \ileyk{The 3D setup obtained through this procedure is then used as an initial state that we now relax in full 3D.} Hereafter, the origin of time is taken at this point and the full 3D setup is run over $\sim 70r_g/c$.

\section{Results}
\label{sec:results}

\subsection{Plasma density and magnetic flux ropes}
\label{sec:n_and_B}

The 3D simulations that we performed closely resemble the 2.5D ones from \pp. The magnetosphere is subdivided in regions presenting different magnetic field topology: (i) open and twisted magnetic field lines threading the \bh event horizon in the polar region, along the \bh spin axis, (ii) open magnetic field lines anchored on the Keplerian disk lying in the equatorial plane, (iii) between the \isco and the separatrix, closed magnetic field loops coupling the disk to the \bh ergosphere and (iv) a cone-shaped current layer where electromagnetic energy is dissipated via magnetic reconnection. The first three regions are essentially force-free, with magnetic field lines rotating at $\sim\omega_{BH}/2$ in the jet and at the Keplerian speed of their footpoint on the disk in the other regions. The Poynting energy fluxes extracting \bh rotational energy in the jet and to the disk are also comparable to the 2.5D results. The 3 force-free regions meet at the Y-ring at the basis of the current layer which supports the discontinuity of the magnetic field components between the jet and the field lines anchored on the disk. The current layer envelops the jets and is a site of vivid magnetic reconnection at the Y-ring but also at transient X-points forming further away. We introduce the azimuthal unit vector $\boldsymbol{\hat{\phi}}$ and the unit vector $\boldsymbol{\hat{l}}$ orthogonal to $\boldsymbol{\hat{\phi}}$, colinear to the intersection between the poloidal plane and the current layer, and pointing away from the \bh. \ileyk{Both vectors are represented in Figure\,\ref{fig:25D}.} The geometry and dimensions of the Y-ring are identical to what had been observed in 2.5D (see Figure 8 in \pp): it lies at $\sim$1.8$r_g$ above the disk and at a distance of $r_{Y,\perp}\sim 4r_g$ from the spin axis (see also Figure\,\ref{fig:triskell}). In Fig\,\ref{fig:plasma_density}, we show a 3D volume rendering of the pair plasma density (corrected for spherical dilution by a factor $r^2$) which peaks in the current layer. The red rings in the right panel stand for the location of the Y-rings where flux ropes grow before eventually detaching and flowing away along the current layer.

\begin{figure}
\centering
\includegraphics[width=0.99\columnwidth,height=7.5cm]{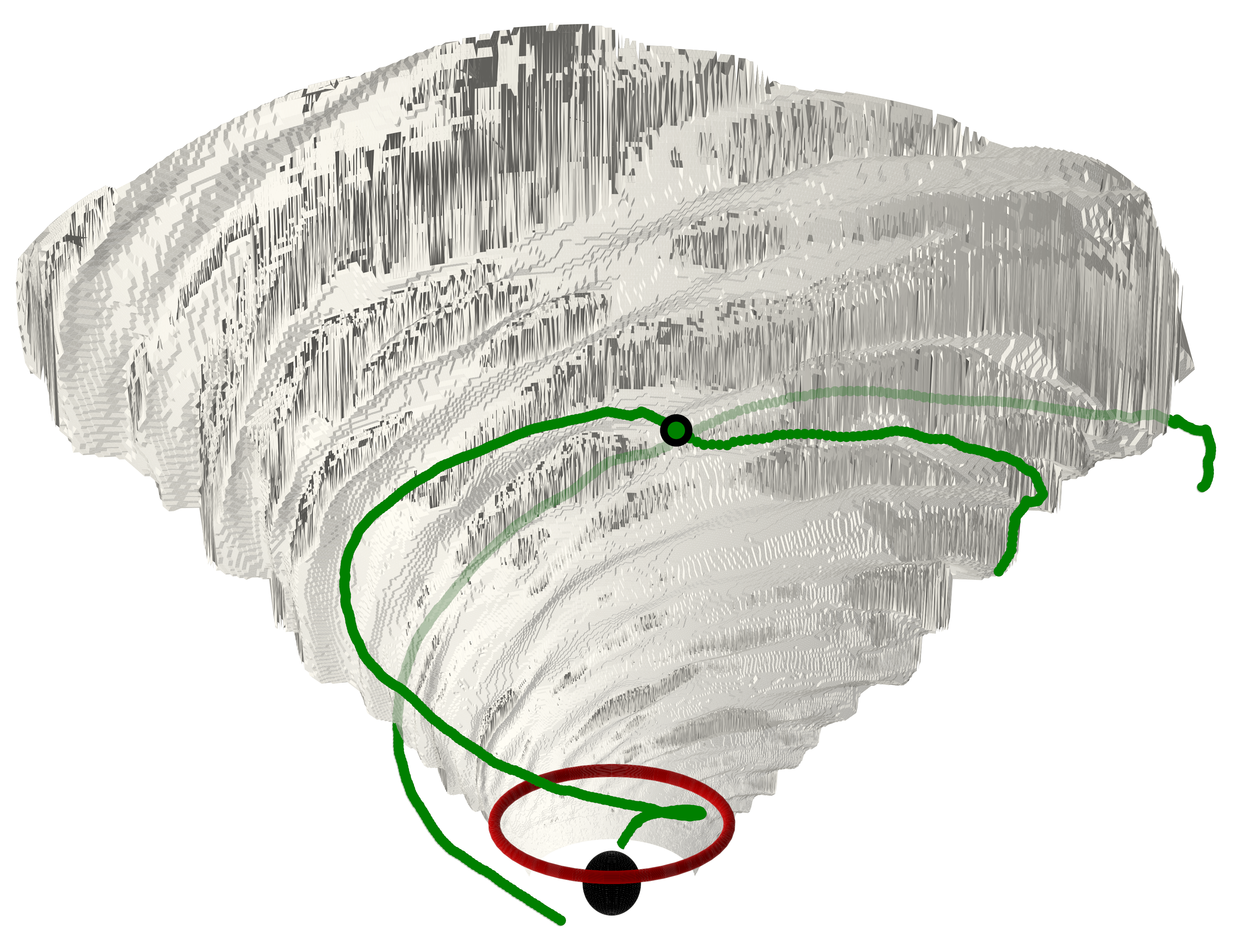}
\caption{Magnetic field lines (solid green) reconnecting across the current layer. The corresponding X-point\ileyk{, where magnetic field lines are anti-parallel,} is the green dot with a black edge. \ileyk{The polarity of the toroidal magnetic field reverses across the white surface.} The red annulus is the Y-ring and the black sphere is the event horizon. }
\label{fig:cs_w_B}
\end{figure} 

A major novelty with respect to 2.5D simulations is the reconnection of the toroidal magnetic component. Both the poloidal and the toroidal components of the field reverse their polarity across the layer, but 2.5D simulations can only capture reconnection in the poloidal plane. They miss the dominant toroidal component. In Fig\,\ref{fig:cs_w_B}, the white surface is the locus of points where the toroidal magnetic field component changes sign \ie where the radial component of the current spikes. It approximately locates the current layer. Along it, the ripples correspond to flux ropes whose combined extension along $\boldsymbol{\hat{l}}$ and $\boldsymbol{\hat{\phi}}$ is well visible. In addition, we represented \ileyk{in green two fiducial magnetic field lines reconnecting at an X-point in the current layer (black-circled green dot). The forefront magnetic field lines threads the event horizon while the background one is anchored on the disk. The strong twisting of these field lines highlights the combined role of the poloidal and toroidal components of the magnetic field}: the anti-parallel reconnection proceeds along \ileyk{obliques in the ($\boldsymbol{\hat{l}}$,$\boldsymbol{\hat{\phi}}$) plane} in the cone-shaped current layer. There is no guide field since the reconnecting layer is not embedded in a background magnetic field transverse to the current layer with a preferential net polarity. Near the separatrix, as the quickly rotating magnetic field lines endure differential torques, they stretch and open up under the action of the tearing instability with a quasi-periodicity of a few $r_g/c$ comparable to what we had found in 2.5D simulations for $a=0.99$. In our 2.5D simulations, the drift-kink instability \citep{Zenitani2007,Barkov2016} stemmed only from the poloidal component of the current along $\boldsymbol{\hat{l}}$ induced by the discontinuity of the toroidal component of the magnetic field across the current layer. Here, it is also driven by the toroidal component of the current associated to the jump of the $\boldsymbol{\hat{l}}$-component of the magnetic field. It further bends the flux ropes and corrugates the current layer but does not inhibit the tearing instability which eventually dominates and regulates the formation of flux ropes. Although the current sheet is broadened, the flux ropes remain confined in a narrow volume around the mid-plane of the cone-shaped current layer (see Fig\,\ref{fig:plasma_density}). 

\subsection{Current layer and reconnection rate}
\label{sec:rec_rate}

\begin{figure}
\centering
\includegraphics[width=0.99\columnwidth]{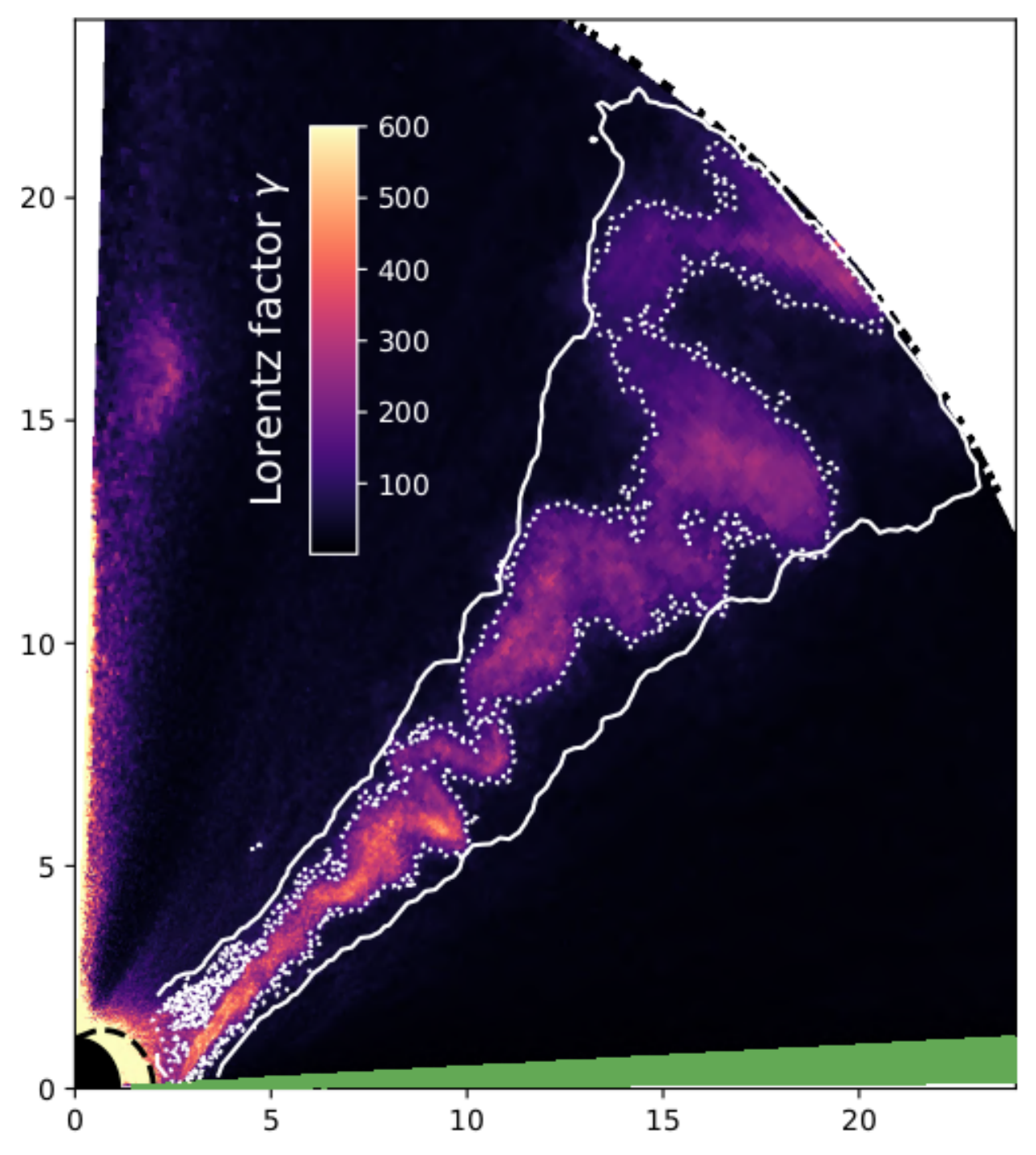}
\caption{Poloidal slice of the mean electron and positron Lorentz factor. The solid white line is a slice of the 3D closed surface encompassing the whole current layer while the dotted line locates, for this slice, the current layer. The black disk at the origin (resp. the black dashed line) is the \bh event horizon (resp. the ergosphere) and the light green region in the equatorial plane stands for the disk.}
\label{fig:lorentz}
\end{figure} 

\begin{figure}
\centering
\includegraphics[width=0.99\columnwidth]{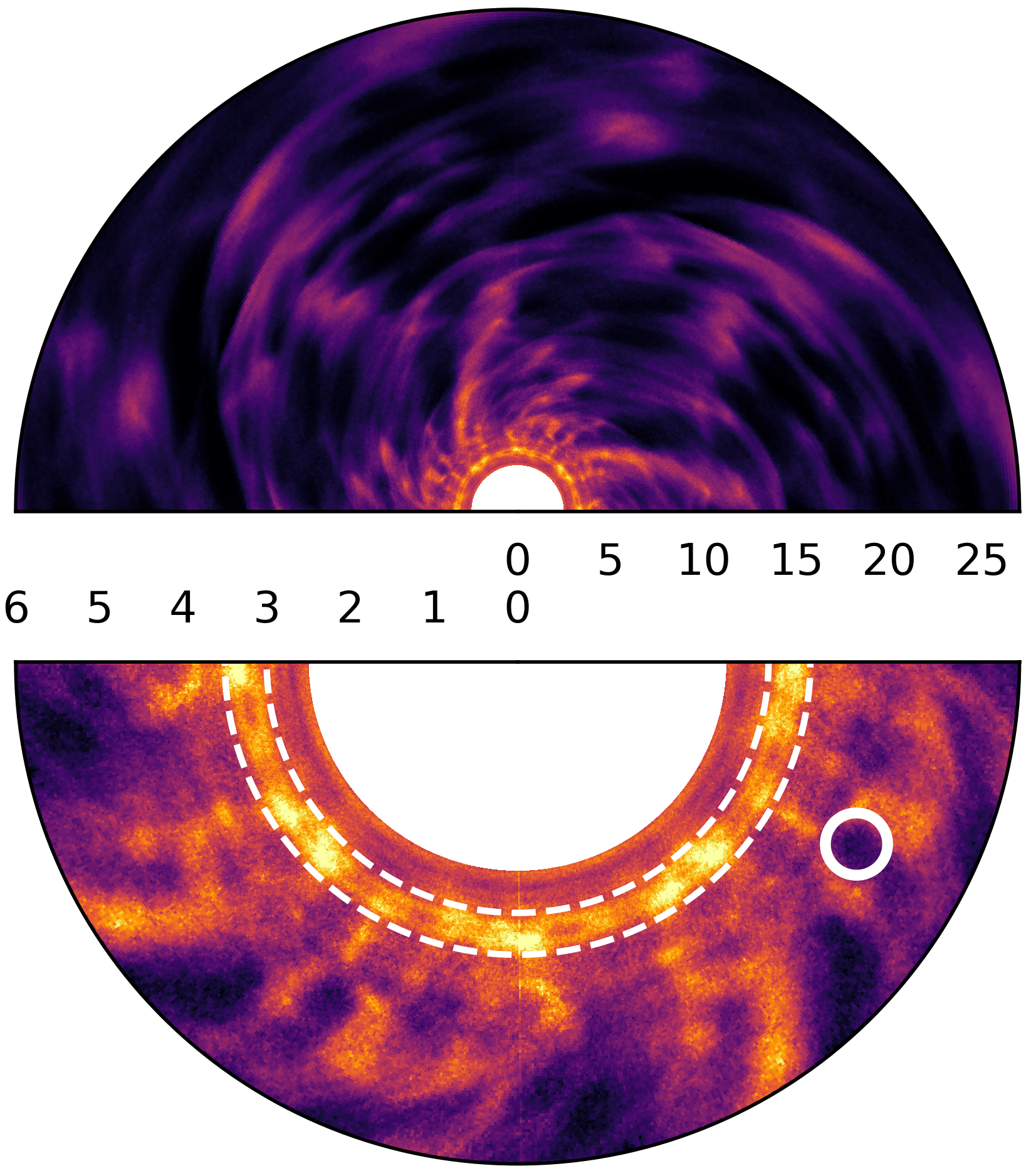}
\caption{Top view projection of the flux ropes in the full and inner current layer (resp. top and bottom panels), with the same color map as in Figure\,\ref{fig:plasma_density}. Numbers indicate the length in $r_g$ along the current layer, with the outer edge of the inner white disk located at the current layer footpoint on the disk $r_{ft}\sim 2.5r_g$. In the bottom panel, the white dashed lines locate the Y-ring. \ileyk{The white circle surrounds an X-point and delimits the region where we compute the transverse velocity profiles} in Fig\,\ref{fig:rec_rate}.}
\label{fig:plasmoids}
\end{figure}

The sign inversion of the toroidal magnetic field represented in Figure\,\ref{fig:cs_w_B} provides a surface for the current layer but it does not grant access to its local thickness. In order to estimate the volume encompassed by the current layer and locate precisely the flux ropes as a function of time, we use the following procedure. The current layer stands out as a region where the particle Lorentz factors are much higher, as visible in the poloidal slice in Figure\,\ref{fig:lorentz}. \ileyk{In the ergosphere, particles show a high Lorentz factor because of our frozen boundary conditions on the equator and because of the ad hoc injection method we rely on. Accounting for pair creation for instance would have probably removed this numerical artifact \citep{Crinquand2020}. Furthermore, although weak in the vicinity of \sgr, synchrotron cooling would be enough to significantly lower the particle Lorentz factor $\gamma$ in force-free regions like the ergosphere, effectively suppressing their contribution to the emission. Near the pole, the high Lorentz factors result from numerical artifacts and should not be taken seriously.} For each slice, we compute the density-weighted Lorentz factor squared which plays the role of a tracer for the current layer
\begin{equation}
\label{eq:tracer}
    \langle \gamma^2 \rangle _n = (n_+\gamma_+^2+n_-\gamma_-^2)/(n_++n_-),
\end{equation}
where $n_+$ (resp. $n_-$) and $\gamma_+$ (resp. $\gamma_-$) stand for the number density and Lorentz factor of the positrons (resp. of the electrons). We later clip out the cells below a discriminating threshold value to be left with the region within the dotted white line in Figure\,\ref{fig:lorentz}. We then integrate the aforementioned tracer in the local transverse direction and over the current layer width to produce a column density representation of the plasma in the current layer such as the one in Figure\,\ref{fig:plasmoids}. \ileyk{With this tracer quantity, the plasma-loaded flux ropes stand out as contrasted structures. As the poloidal bulk velocity of the flux ropes increases from close to 0 at the Y-ring to mildly relativistic speeds, the apparent shape of the flux ropes is flattened by length contraction (see also Section\,\ref{sec:099}).} 

\begin{figure}
\centering
\includegraphics[width=0.99\columnwidth]{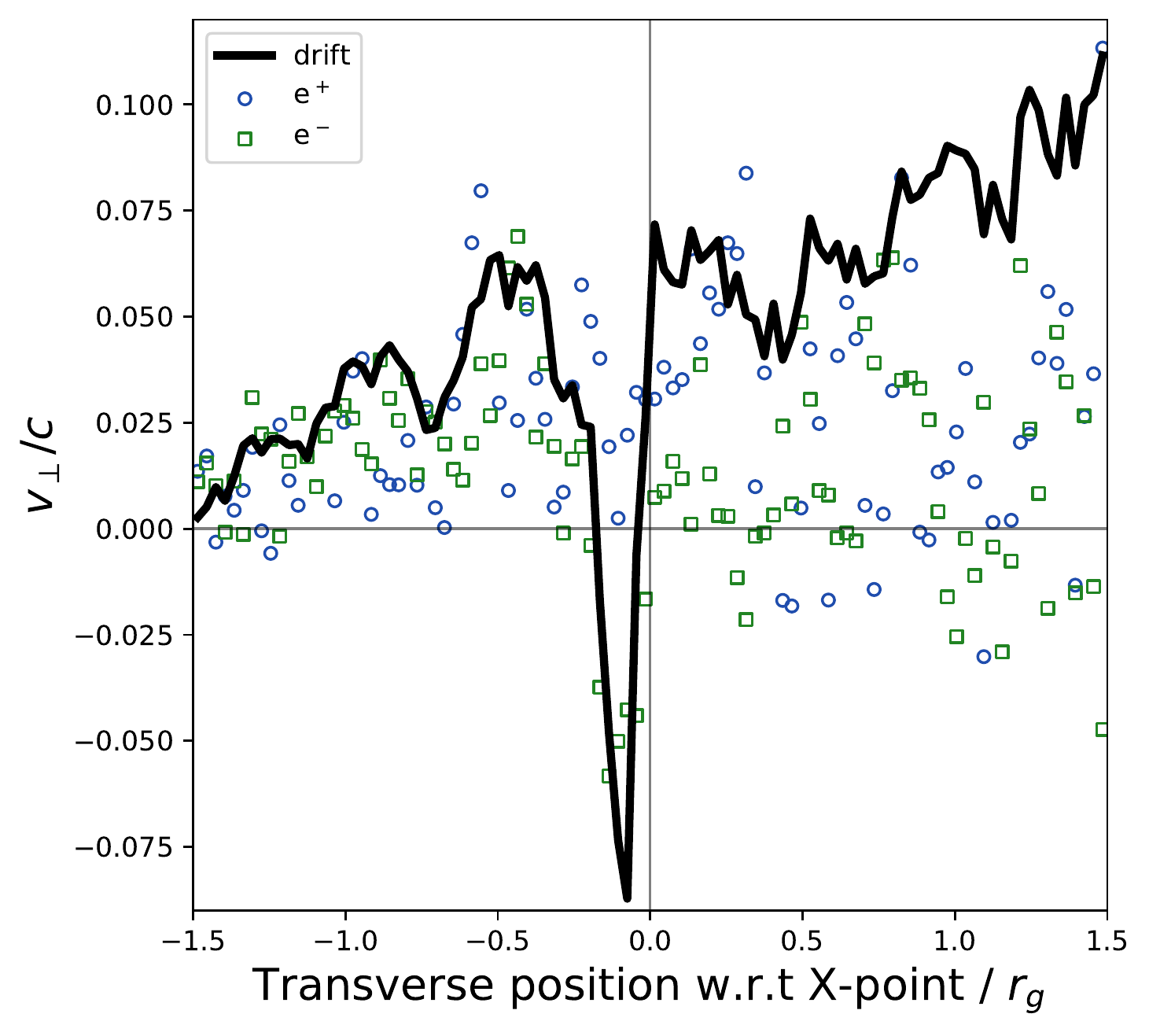}
\caption{Profiles along a line transverse to the current layer and passing through a X-point. We show the normal components of the drift velocity (black solid line) and of the positron and electron velocities (respectively blue circles and green squares).}
\label{fig:rec_rate}
\end{figure}

Contrary to a simple slice (at constant $\theta$ for instance), this method enables us to follow all the flux ropes, including those which momentarily leave the current layer mid-plane under the action of the kink instability. It also captures the X-points, where the magnetic field reconnects and where the width of the current layer reaches its minimal thickness set by the plasma skin depth. They manifest in the column density representation as expanding regions of tracer depletion. For instance, in Figure\,\ref{fig:plasmoids}, a representative X-point in a fiducial snapshot is shown with a white circle in the bottom panel. At this X-point, we measure a current layer thickness of $\delta\sim 0.1 r_g$, resolved with 3 to 5 grid points. It is consistent with the number of flux ropes which form at the Y-ring, typically 6 per $\pi/2$ azimuthal quarter (see bright spots at the Y-ring in Figure\,\ref{fig:plasmoids}). Indeed, the maximally unstable mode of the tearing instability yields an azimuthal distance between two successive flux ropes of $\lambda=2\pi\sqrt{3}\delta$ \citep{Zelenyi1979,Zenitani2007,Cerutti2014} and thus an initial number of flux ropes per quarter formed:
\begin{equation}
\label{eq:N0}
    N_0=\frac{\pi r_{Y,\perp}}{2\lambda}= \frac{r_{Y,\perp}}{4\sqrt{3}\delta}\sim 6 \left(\frac{r_{Y,\perp}}{4r_g}\right) \left(\frac{\delta}{0.1r_g}\right)^{-1}
\end{equation} 
where $r_{Y,\perp}\sim 4r_g$ is the radius of the Y-ring (\ie the distance of the Y-points to the spin axis, see also Figure\,\ref{fig:triskell}). \ileyk{Since $N_0>1$, it justifies a posteriori our choice of working only over the range $\phi=0$ to $\phi=\pi/2$.} We estimate the reconnection rate $\beta_{\rm rec}$ based on the upstream velocities near the aforementioned fiducial X-point. As visible in Figure\,\ref{fig:plasmoids}, this X-point is approximately located 1$r_g$ beyond the Y-ring in the current layer (white dashed circles). In order to diminish the numerical noise, we work with an averaging box which is 3$r_g$ long in the transverse direction ($x$ axis in Figure\,\ref{fig:rec_rate}) and 0.2$r_g$ wide in the two directions $\boldsymbol{\hat{l}}$ and $\boldsymbol{\hat{\phi}}$ locally coplanar to the current layer such as the averaging box covers the dark region surrounding the white spot in Figure\,\ref{fig:plasmoids}. We thus obtain the transverse profiles of the component of the particle bulk velocities normal to the current layer, $v_{\perp}$, displayed in Figure\,\ref{fig:rec_rate}. The sign convention is such that $v_{\perp}>0$ (resp. $v_{\perp}<0$) if particles are above the current layer and move towards it or if they are below the current layer and move away from it (resp. if they are above the current layer and move away from it or if they are below the current layer and move towards it). Electron and positron velocities are both very noisy, although electrons (green squares) show a sharp drop correlated with the one in the $\mathbf{E}\times\mathbf{B}$-drift velocity profile (black solid line) computed from the electromagnetic fields on the grid. At this X-point where the magnetization $\sigma \gg 1$ and where the Alfven speed is close to the speed of light, we thus measure a transverse velocity step of $\Delta v_{\perp}\sim 0.15c$ corresponding to a magnetic reconnection rate $\beta_{\rm rec}=\Delta v_{\perp}/(2c)\sim 7\%$. It must be understood as a lower limit since we rely on an approximate localization method for the X-point.

This high reconnection rate is typical of what is found from PIC simulations of magnetic reconnection of collisionless pair plasma without guide field in the relativistic regime \citep{Cassak2017,Werner2018,Crinquand2021,Goodbred2023}. Surprisingly, although we also capture the reconnection of the toroidal component, this rate is close from what we had obtained in 2D ($\beta_{\rm rec}\sim 5\%$) where only the poloidal component along the current layer was reconnecting. We think it could be due to the broadening of the current layer induced by enhanced turbulence in 3D, partly caused by the capacity of the drift-kink instability to grow in both the $\boldsymbol{\hat{l}}$ and $\boldsymbol{\hat{\phi}}$ directions \ileyk{\citep{Sironi2014,Guo2015,Werner2017}}. The current layer is less laminar than in 2D and we measure a thickness of $\delta\sim 0.1 r_g$, twice thicker than what we had obtained in 2D simulations. Lower reconnection rates in 3D have been commonly reported \citep{Sironi2014,Zhang2021,Werner2021} and ascribed to magnetic flux diffusion and topological alternatives to reconnection which do not exist in 2D: although the amount of magnetic flux available for reconnection is higher, part of it is annihilated before reconnecting, yielding a net reconnection rate comparable or even lower than in 2D \citep{Werner2021}. 




\section{Discussion}
\label{sec:discussion}

\subsection{Flux ropes kinematics}
\label{sec:kinematics}

\subsubsection{For a \bh spin $a=0.99$}
\label{sec:099}

\begin{figure}
\centering
\includegraphics[width=0.99\columnwidth]{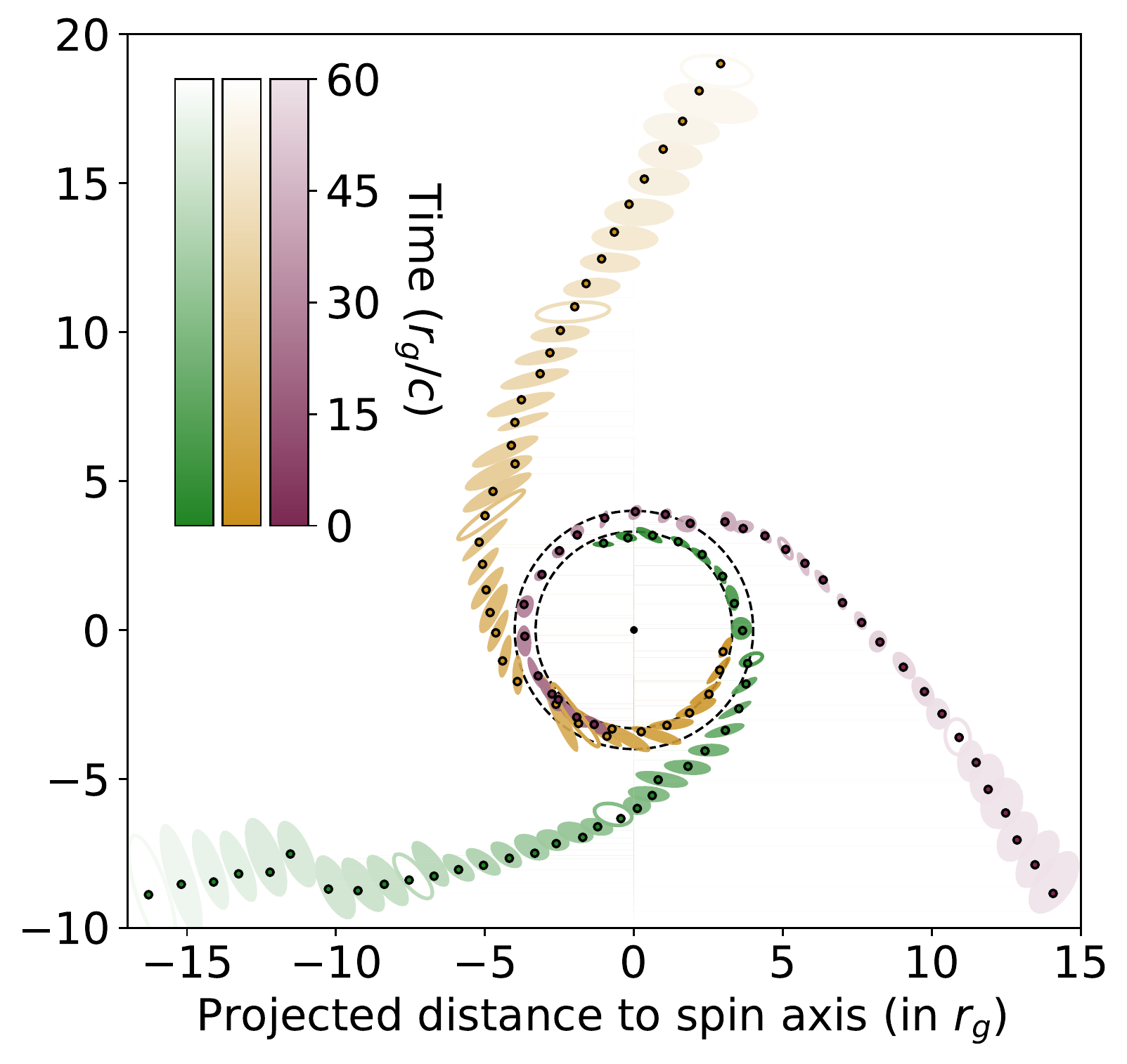}
\caption{Trajectory of 3 plasma-loaded flux ropes projected in the equatorial plane (top view). The dots indicate the center of the flux rope while the ellipses represent their aspect ratio, orientation and spatial extension. The 2 concentric dashed circles stand for the Y-ring. The color is the time lapse from departure for each flux rope and empty ellipses are shown every 15$r_g/c$.}
\label{fig:triskell}
\end{figure}

\begin{figure}
\centering
   \includegraphics[width=0.98\columnwidth]{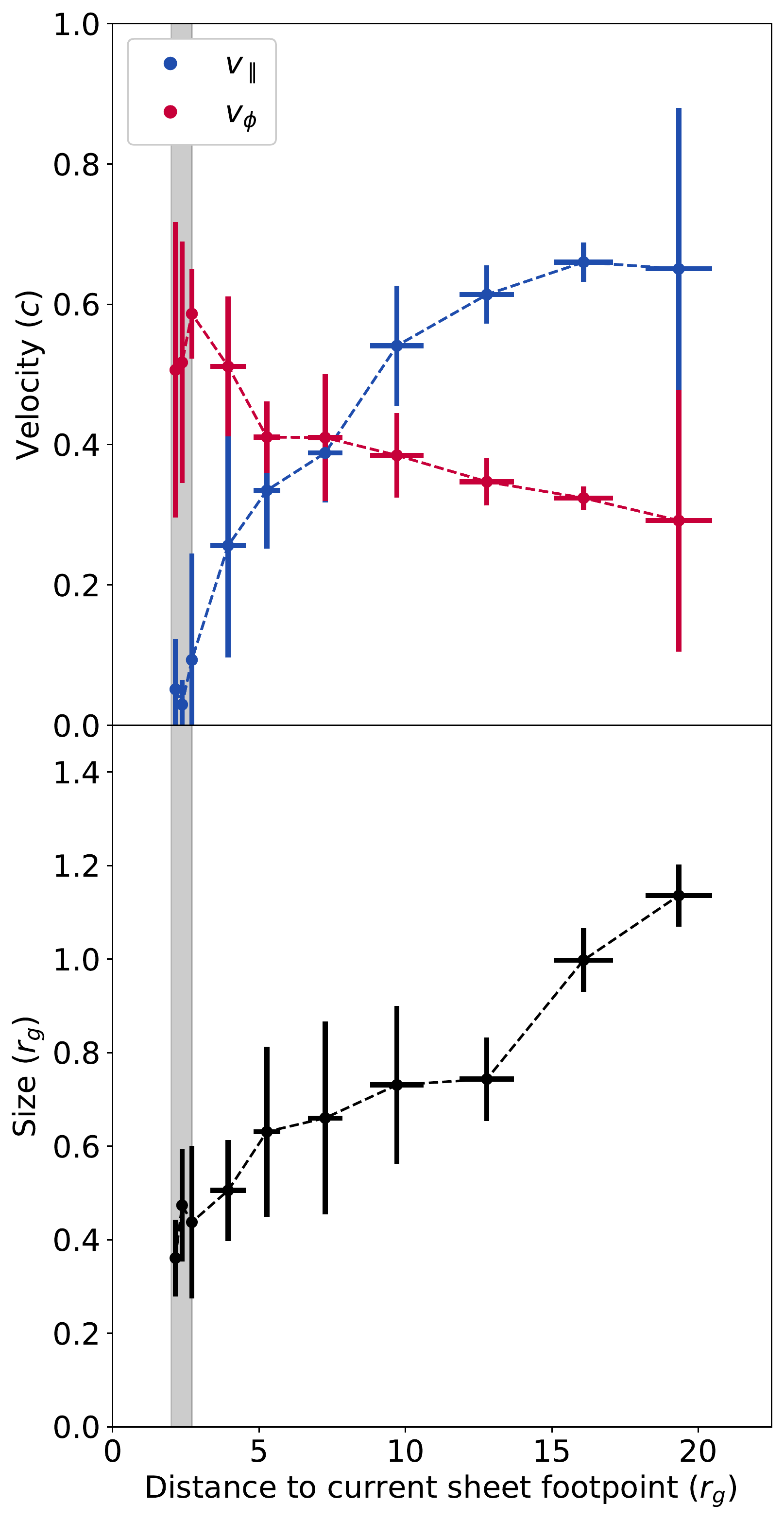}
\caption{\textit{(top)} Poloidal (in blue) and azimuthal (red) velocity components of the flux ropes in the current layer as a function of the distance to the current layer footpoint on the disk mid-plane. \textit{(bottom)} \ileyk{Effective radius} of the flux ropes as they flow away from the Y-ring. The inner grey shaded region stands for the Y-ring.}
\label{fig:kinematics}
\end{figure}

We now proceed to characterize the trajectory of the dense plasma contained in the flux ropes in the current layer. Owing to the low viewing angle of \sgr \citep{Jimenez-Rosales2020,GravityCollaboration2021,Collaboration2022,Collaboration2022,Wielgus2022}, we focus on the apparent motion of the flux ropes as seen face-on. We use the tracer introduced in Equation\,\eqref{eq:tracer} in order to identify 3 major flux ropes and measure their position in the plane of the sky every $\sim 1.6r_g/c$. In Figure\,\ref{fig:triskell}, the dots trace the \ileyk{projected positions} of the flux ropes as a function of time. For the 3 trajectories, a $\pm 2\pi/3$ shift rotation around the \bh spin axis was included for the sake of visualization, and the origin of time is set at the first snapshot where the flux rope was identified. Each trajectory lasts approximately 50$r_g/c$ between the Y-ring (black dashed circles) and the outer extent of the simulation space, and it can be subdivided in two parts: (i) the loading phase, when the flux rope is still attached to the separatrix and progressively grows as the separatrix stretches and (ii) the outflowing expansion phase, once the flux rope detaches from the separatrix due to the tearing instability and flows away along the current layer. In Figure\,\ref{fig:kinematics}, we report reduced kinematic data concerning the velocity profiles in the current layer in the $\boldsymbol{\hat{l}}$ and $\boldsymbol{\hat{\phi}}$ directions (resp. $v_{\parallel}$ in blue and $v_{\phi}$ in red). The two vertical dashed lines show the extremal positions of the outermost point of the separatrix as the separatrix stretches out. These velocity profiles have been binned and obtained from averaging the values of the 3 trajectories represented in Figure\,\ref{fig:triskell}. The large uncertainties on $v_{\parallel}$ and $v_{\phi}$ are due to the inherent inaccuracies of the method we use to locate the centers of the flux ropes: flux ropes merge as they flow away and since they are elongated filaments, their center is often ill-defined, hence the glitches in the trajectories display in Figure\,\ref{fig:triskell} which translates in sudden jumps in the velocity profiles and, once binned, in large error bars in Figure\,\ref{fig:kinematics}.

The duration of the loading phase is set by the quasi-periodic breathing motion of the extremity of the separatrix as the separatrix stretches out and opens up. We measure a quasi-periodicity of $T_Y\sim 10-20 r_g/c$, somewhat longer than the $7r_g/c$ we had measured in 2.5D for $a=0.99$ (see Figure 7 in \pp). This increased duration might be the result of the non-axisymmetric toroidal magnetic tension whose progressive growth and sudden release contributes to the mechanism. During this lapse of time $T_Y$, the flux ropes rotate around the spin axis with a speed $v_{\phi,Y}\sim 0.5-0.6c$ (see red data between the vertical dashed lines in Figure\,\ref{fig:kinematics}) and with an apparent orbital period of $T_{orb,Y}\sim 45 r_g/c$. We qualify this period of "apparent" in order to stress that seen face-on, it can be confused with an orbital motion in the equatorial plane of the \bh. However, in this model, the flux rope trajectory during the loading phase is set by its connection with the separatrix and it takes place along the Y-ring located above the equatorial plane. Also, we notice that the flux rope rotation speed $v_{\phi,Y}$ is $\sim 30\%$ lower than the Keplerian speed of the separatrix footpoint at $r_{ft}\sim 2.5 r_g$ on the disk. We interpret this discrepancy as the outcome of the progressive stretching of the separatrix along the $\boldsymbol{\hat{l}}$ and $\boldsymbol{\hat{\phi}}$ directions: \ileyk{near the Y-ring, the inertia of the plasma becomes non-negligible and the force-free approximation breaks down. Consequently, at the Y-ring, the separatrix rotates slower than at its footpoint, the toroidal component grows, the separatrix inflates and eventually, it opens up.} Still, this motion looks super-Keplerian compared to what would be expected for a hot spot on a Keplerian orbit in the plane of the sky at a distance $r=r_{Y,\perp}$ from the \bh, which would yield $v_{\phi}\sim 0.4-0.5c$. We can deduce from the duration of the loading phase $T_Y$ and from the apparent orbital period at the Y-ring $T_{orb,Y}$ the typical fraction of circle which is spanned by the flux ropes before detaching from the Y-ring:
\begin{equation}
    \frac{T_Y}{T_{orb,Y}} \sim 30-40\%
\end{equation}
in agreement with Figure\,\ref{fig:triskell}. 

Once the flux rope enters the outflowing expansion phase, it describes a spiral trajectory of increasing pitch angle such that beyond $r_g$, the $\boldsymbol{\hat{l}}$-component is higher than the azimuthal component of the bulk velocity (Figure\,\ref{fig:kinematics}). \ileyk{Toward the outer edge of the simulation space, the trajectory is almost radial as seen face-on.} Interestingly enough, the azimuthal speed profile beyond the Y-ring decreases slower than what we would have expected purely from angular momentum conservation, probably due to the interplay of the flux ropes with its immediate environment. The plasma within the flux rope experiences magnetic confinement and exerts kinetic pressure on the surrounding magnetic field lines. In parallel, we measure a progressive increase of the outward bulk speed $v_{\parallel}$ due to the magnetic slingshot mechanism associated to the opening of the separatrix: once the flux rope detaches from the Y-ring due to the tearing instability, it no longer enjoys the retaining magnetic tension from the separatrix so its super-Keplerian azimuthal speed leads to an increase of the speed along $\boldsymbol{\hat{l}}$. Similar velocity profiles are obtained for pulsars beyond the light cylinder when working with a split monopole: over $5-6$ light cylinder radii, the azimuthal (resp. radial) component of the velocity decreases slower (resp. increases slower) than with a pure monopole \citep{Cerutti2017a}. Also, the $v_{\phi}$ profile is similar to the one of ejected flux ropes in the polar regions of GRMHD simulations obtained by \cite{Nathanail2020}, but our $v_{\parallel}$ increases very fast compared to their radial velocity profile. Consequently, our trajectories are more open.


Finally, we find that, between the beginning of the loading phase up to the regime when the $\boldsymbol{\hat{l}}$-component of the velocity dominates, the trajectory of the flux rope seen face-on covers an azimuthal extent of $4\pi/3$ to $3\pi/2$, but it never loops over its initial azimuthal position. It can be seen as a consequence of the fact that whatever the spin, the tearing instability acts on a shorter timescale than the apparent orbital period at the Y-ring. The shape of the trajectories we find is in qualitative agreement with the GRAVITY hot spot observations. We predict that if a single macroscopic flux rope is responsible for one observed hot spot, any centroid shift motion will yield an incomplete projected trajectory. Until now, all the centroid shifts captured by the GRAVITY collaboration during flares fulfill this requirement \citep{Abuter2018b}.

\subsubsection{Prediction for the spin of \sgr}
\label{sec:lower_spin}

\begin{figure}
\centering
\includegraphics[width=0.99\columnwidth]{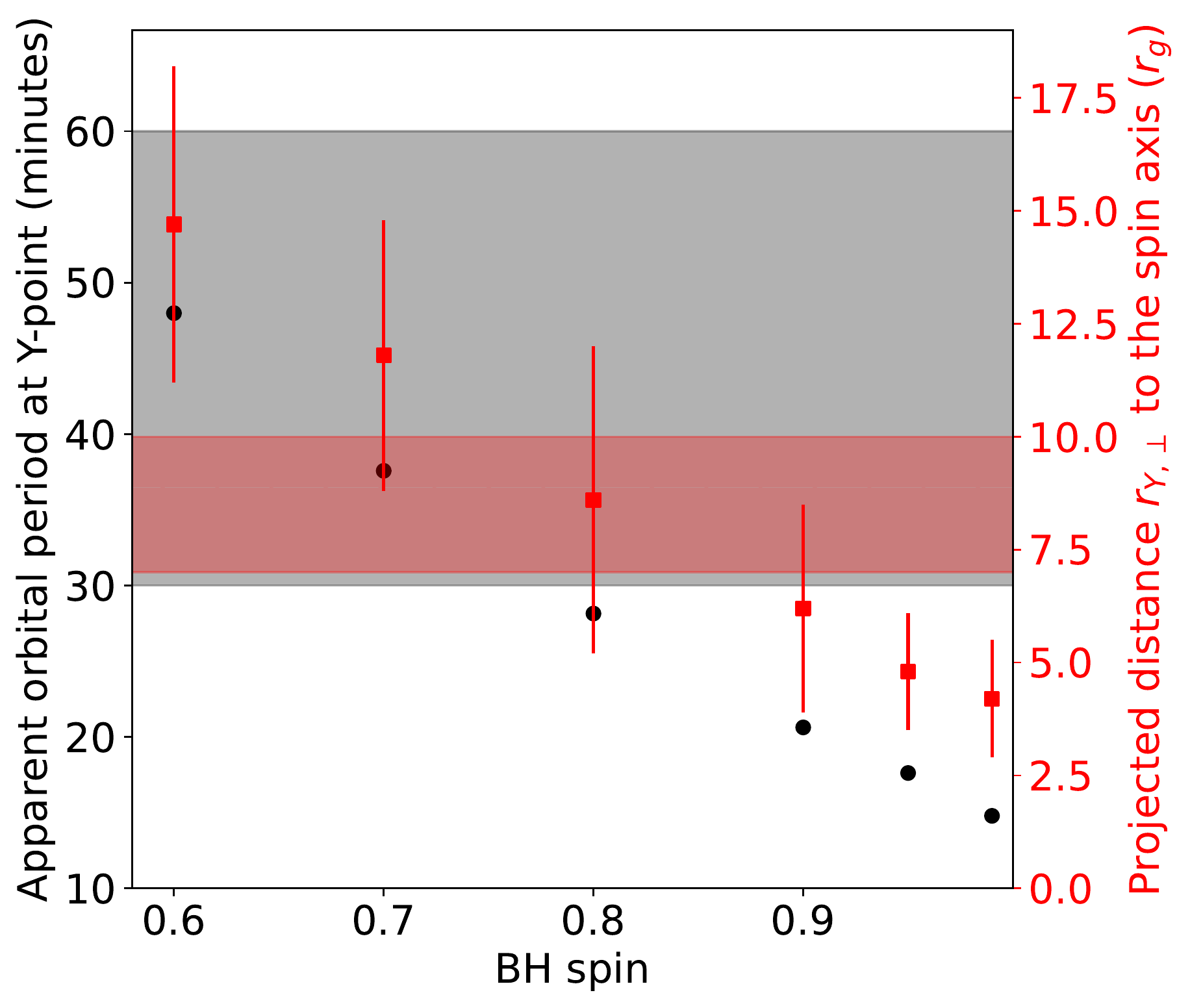}
\caption{Apparent orbital period at the Y-ring (left axis, black dots) and projected distance of the Y-ring to the spin axis (right axis, red squares) as a function of the \bh spin. The points for \bh spin $a<0.99$ are extrapolated from the 2.5D simulations of \pp but include the correction factor on the plasma velocity at the Y-ring we measure in 3D simulations. The hatched regions in black and red stand for the constraints set by the GRAVITY observations and modeling of the 2018 hot spots \citep{Baubock2020}.}
\label{fig:GRAVITY}
\end{figure}

We now confront to observations our results on the flux rope kinematics that we extrapolate to lower \bh spin based on the 2.5D results we obtained in \pp, and on the similarities and differences we observed in 3D for a spin of $a=0.99$. Most importantly, GRAVITY observations highlight two major constrains on the motion of the hot spots: their apparent orbital period, $T_{orb,Y}\sim 30-60$ minutes and their distance to the spin axis $r_{Y,\perp}\sim 7-10r_g$. Assuming most of the emission would originate from flux ropes before they detach from the Y-ring, we can compare these values to the ones obtained for lower \bh spins. In Figure\,\ref{fig:GRAVITY}, we plotted the apparent orbital period (in black) and the projected distances of the Y-ring to the spin axis (in red). \ileyk{The apparent orbital period is based on the location of the separatrix' footpoint and accounts for the aforementioned $30\%$ lower speed at the Y-ring with respect to the Keplerian speed at the footpoint (due to the non-negligible inertia of the plasma near the Y-ring).} The large error bars on the radius of the Y-ring correspond to its upper and lower values, and it is representative of the breathing motion of the separatrix which shows a larger amplitude for lower \bh spins. 

For a spin of $a=0.99$, the evolution is too fast and confined to a region too close from the spin axis to match the observational constraints (hatched regions). On the reverse, for a \bh spin $a<0.65$, the Y-ring is located too far from the spin axis. Since the flux ropes flow outwards in the current layer and that the region encompassed by closed magnetic field lines coupling the \bh to the disk is essentially force-free, no hot spot could be produced close enough from the spin axis through this mechanism. The constraint on the distance to the spin axis suggests a \bh spin ranging between $0.65$ and $0.9$, while the timing of the hot spot indicates a \bh spin lower than $0.8$ to reproduce the apparent orbital period. If the hot spot is indeed associated to the formation of flux ropes in a cone-shaped current layer, our analysis favors a \bh spin of $0.65$ to $0.8$ for \sgr.

\ileyk{We had noticed in \pp that the quasi-periodic breathing motion of the extremity of the separatrix was twice slower in the $a=0.6$ case compared to the $a=0.99$ case.} If this ratio still holds in 3D, it would mean that the fraction of the Y-ring spanned by the flux ropes before they detach decreases for lower spins, down to:
\begin{equation}
    \frac{T_Y}{T_{orb,Y}} \sim 20-25\%.
\end{equation}
for $a=0.6$ \ileyk{because $T_Y$ increases slower with spin than $T_{orb,Y}$}. Since we do not expect the dynamics during the outflowing expansion phase to significantly depend on the \bh spin, it would overall lead to more open trajectories than in Figure\,\ref{fig:triskell}. \ileyk{Consequently, if a single plasmoid formed and ejected from the Y-point is responsible for a full NIR flare in \sgr, we expect the trajectories observed with astrometry to be necessarily open.}



\subsection{Flux rope growth}
\label{sec:growth}

Before moving on to the synchrotron radiation emitted by each of the 3 flux ropes from Figure\,\ref{fig:triskell}, we evaluate their size along their trajectory. \ileyk{Every 1.6$r_g/c$, we fit the projected shape of the flux rope with an ellipse (see Figure\,\ref{fig:triskell}) and compute the radius of the circle with same surface (hereafter the effective radius, black dots in Figure\,\ref{fig:kinematics}).} The aim of the procedure is to compute synchrotron power and extract trends in the evolution of the flux ropes' size, not to be used as an accurate prediction. A power law fit of the flux ropes' \ileyk{effective radius} as a function of the distance to the Y-ring gives an exponent of $\sim 0.6$, in agreement with the mean radial profile of the plasma density which goes as $r^{-1.8}$. It indicates that as flux ropes flow away from the Y-ring, they grow slower than the distance between them does.

Increasing confidence in favor of flux ropes growth in current layers through coalescence up to macroscopic scales has built up in the community within the last decade \citep{Loureiro2012a,Sironi2016a,Philippov2019,Cerutti2021}. Models of hierarchical merging flux ropes have emerged and indicate that in highly magnetized collisionless plasmas such as \bh magnetospheres, giant flux ropes can form within a few light crossing time \citep{Zhou2019,Zhou2020,Zhou2021}. In our simulations, while flux ropes are progressively loaded with plasma at the Y-ring before they detach, their later growth during the outflowing expansion phase is dominated by successive mergers between each other. The initial fragmentation of the current layer induced by the tearing instability yields a characteristics number of flux ropes which strongly depends on the thickness of the current layer. In Section\,\ref{sec:rec_rate}, we derived in Equation\eqref{eq:N0} an initial number of flux ropes based on the thickness of the current layer we could resolve. However, in realistic conditions, the scale separation between the plasma skin depth and the gravitational radius is orders of magnitude larger than what is achievable in a PIC simulation, a fortiori if the simulation is global and three-dimensional. For a plasma density of $n\sim 10^6$cm$^{-3}$ \citep{GravityCollaboration2021}, we obtain a skin depth $\sim 10^{-9}r_g$ and the current layer is thus much thinner than in our simulations. In these conditions, the eigenmode of the tearing instability would lead to the formation of $\sim 10^8$ flux ropes at the Y-ring. However, mergers between flux ropes initially occur on a timescale $\tau_0$ given by the initial half-separation $d_0$ between flux ropes:
\begin{equation}
    \tau_0=\frac{d_0}{\beta_{\rm rec} c} \ll \frac{r_g}{c}
\end{equation}
which is very short due to the high reconnection rate and to the small distance between flux ropes. The remaining number $N$ of flux ropes over $2\pi$ after a time $t \gtrapprox r_g/c$ no longer depends on the initial number $N_0$ of flux ropes and is given by \citep{Cerutti2021}:
\begin{equation}
    \label{eq:N_growth}
    N\sim N_0 \frac{\tau_0}{t}\sim \frac{\pi}{\beta_{\rm rec}}\frac{r_{Y,\perp}}{ct}\sim 30 \left(\frac{r_{Y,\perp}/c}{t}\right)\left(\frac{0.1}{\beta_{\rm rec}}\right)
\end{equation} 
with $r_{Y,\perp}\sim 4 r_g$ (for $a=0.99$). The number of flux ropes left after a light crossing time of the Y-ring radius $t=r_{Y,\perp}/c$ is independent of the \bh spin and much smaller than the $10^8$ flux ropes initially formed. Very quickly, the flux ropes are thus bound to coalesce within a few macroscopic structures whose number is only governed by the reconnection rate, which brings support to the approach we undertake of monitoring the evolution of a few giant flux ropes. The good agreement between the number of flux ropes which form at the Y-ring in our simulations (equation\,\ref{eq:N0}) and the analytic estimate\,\eqref{eq:N_growth} brings further support that the skin depth we work with is small enough to accurately capture the macroscopic flux ropes.


\subsection{Synchrotron emission}
\label{sec:synchrotron}

\subsubsection{Contribution per flux rope}

\begin{figure}
\centering
   \includegraphics[width=0.98\columnwidth]{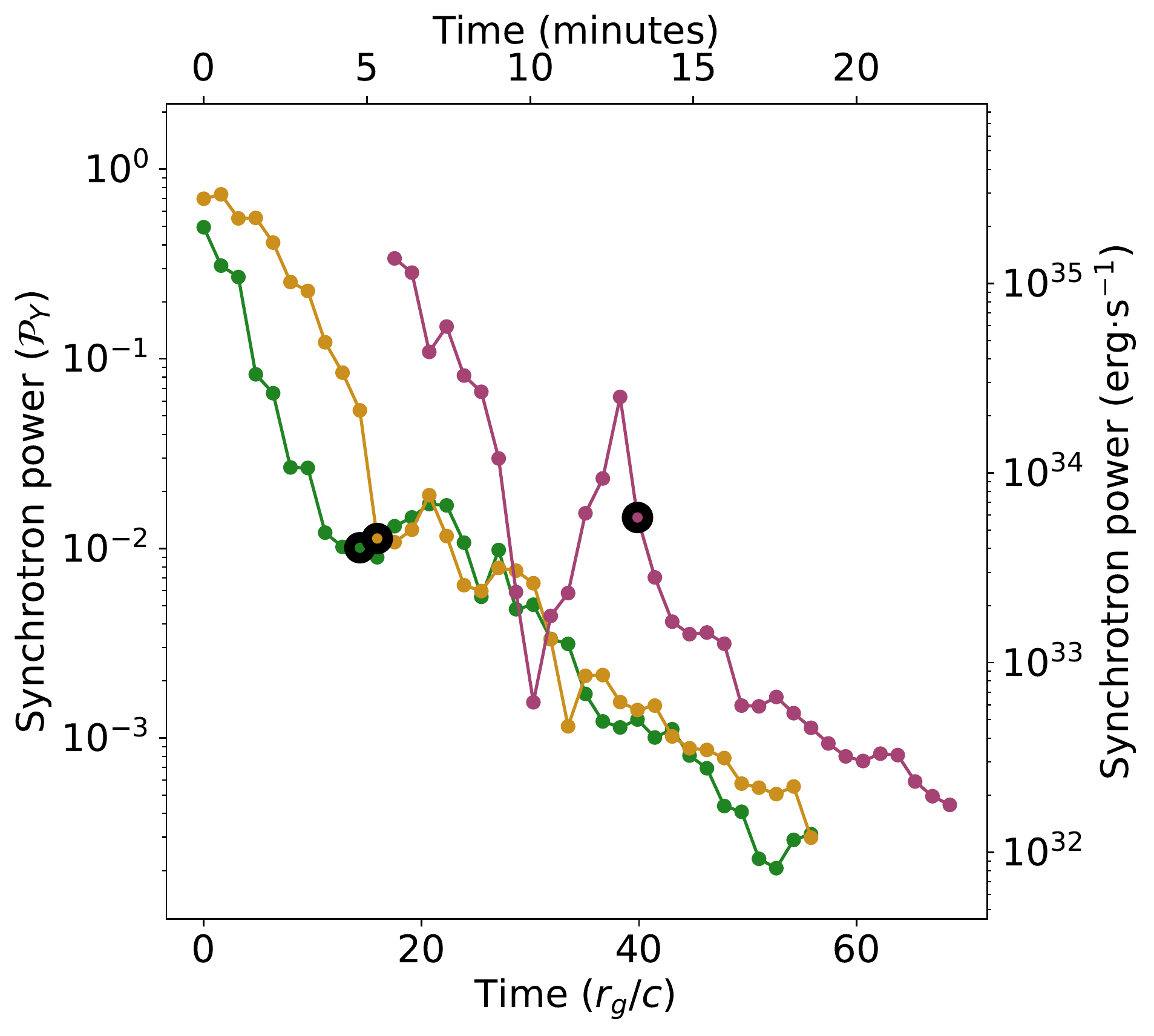}
\caption{\ileyk{Synchrotron power (in units of $\mathcal{P}_Y$ on the left and in erg$\cdot$s$^{-1}$ on the right) from the 3 flux ropes whose trajectory is represented in Figure\,\ref{fig:triskell}, as a function of time (in $r_g/c$ at the bottom and in minutes at the top). All quantities were scaled to \sgr.} Each black circle indicates the approximate time when the flux rope detaches from the Y-ring.}
\label{fig:sync_lc}
\end{figure}

\ileyk{The electrons and positrons reach relativistic Lorentz factors of $\gamma\sim 100-1,000$ through magnetic reconnection in the current layer. The maximum Lorentz factors the particles reach in our simulation are consistent with the magnetization $\sigma\sim 1,000$ we work with.} As they gyrate around the magnetic field lines, particles emit synchrotron radiation with a power per unit volume which is given by:
\begin{equation}
\label{eq:psyn}
    \frac{\mathrm{d}\mathcal{P}_{syn}}{\mathrm{d}V}=\frac{4}{3}\sigma_T c (n_+ \gamma_+^2 + n_- \gamma_-^2) B^2,
\end{equation}
in the limit of isotropic pitch angles and relativistic Lorentz factors, and with $\sigma_T$ the cross section for Thomson scattering. Owing to its large magnetic energy and density, the plasma contained in flux ropes is a source of high synchrotron emissivity. If the \ileyk{magnetosphere} is optically thin to synchrotron radiation, we can integrate this quantity\,Eq.~\eqref{eq:psyn} over the flux ropes to obtain the synchrotron power emitted by each structure as a function of time, represented in Figure\,\ref{fig:sync_lc}. Each color refers to one of the 3 flux ropes displayed in Figure\,\ref{fig:triskell}. The black circles correspond to the moment when each flux rope detaches and transits from the loading phase to the outflowing expansion phase.The glitch during the loading phase of the purple curve, around $t\sim 30 r_g/c$, can be ascribed to the decrease of the flux rope size. It is a manifestation of the large dispersion of the flux rope sizes at the Y-ring, visible in Figure\,\ref{fig:sync_lc}, and it is probably due to our approximate fitting procedure. The main physical feature of interest is the global decay of each curve, especially during the outflowing expansion phase. Since the number of emitting electrons and positrons and their Lorentz factor remain fairly constant over time, it is mostly due to the lack of cooling and quickly decreasing magnitude of the magnetic field with the distance to the \bh. Synchrotron emission peaks at the basis of the trajectory, near the Y-ring, and it decreases by 3 orders of magnitude within 60$r_g/c$.

\ileyk{The normalization of the synchrotron power $\mathcal{P}_Y$ we work with is:
\begin{equation}
    \mathcal{P}_Y=\frac{4}{3}\sigma_T c \sigma_Y^2 n_Y B_Y^2 r_g^3
\end{equation}
where $n_Y$ and $B_Y$ are the density and magnetic field at the Y-ring. A magnetization $\sigma_Y=300$ is introduced to represent the typical Lorentz factor the particles reach in our simulation. The mass of the \smbh \sgr is constrained accurately enough that even with the strong dependence of $\mathcal{P}_Y$ on $r_g$, it remains a minor source of uncertainty for the synchrotron power. Instead, the scale of the synchrotron power depends essentially on the magnetic field and plasma density at the Y-ring. \cite{GravityCollaboration2021} measured the magnetic field in the vicinity of \sgr by fitting multiwavelenght spectra with a uniform synchrotron emitting sphere of $\sim 20 r_g$ \citep[see also][]{Dallilar2022}. Methods based on spectral fits can only measure the magnetic field in regions where accelerated particles radiate, while the innermost regions are devoid of such particles because of cooling and lack of energy source. The values derived from one-zone model may thus be representative of the values near the Y-ring, where most of the emission comes from whatever the \bh spin and wherever the Y-ring. Consequently, we take $B_Y=30$G and $n_Y=10^6$cm$^{-3}$, in agreement with the values reported in the literature \cite[\eg by][]{Bower2019,GravityCollaboration2021,EventHorizonTelescopeCollaboration2022}. The synchrotron power in physical units is displayed in the right $y$-axis in Figure\,\ref{fig:sync_lc}.} \ileyk{The peak luminosity is of the order of $10^{35}$erg$\cdot$s$^{-1}$, which matches a typical NIR flare in \sgr \citep{Ponti2017}.} For electrons and positrons with a Lorentz factor of $\gamma=100-1,000$ in a magnetic field of $B=100G$, the corresponding frequency range is: 
\begin{equation}
\nu=4\cdot 10^{12-14} \left( \frac{\gamma}{100-1,000} \right)^{2} \left( \frac{B}{100\text{G}} \right) \text{Hz}     
\end{equation}
corresponding to a wavelength of 0.75$\mu$m to 75$\mu$m that is to say the infrared band where the flares of \sgr are observed \citep{Genzel2003}. \ileyk{The magnetosphere of \sgr might be even more magnetized than what we assumed but this result shows that $\sigma\sim 1,000$ is a lower limit in order to accelerate particles to Lorentz factors high enough for NIR synchrotron flares.} Energetically speaking, the synchrotron power emitted by one single flux rope is thus fully able to reproduce the observed NIR peak flares but it cannot be sustained for the observed amount of time: over the $\sim 30$ to $60$ minutes (\ie 90-180$r_g/c$) duration of the GRAVITY flares, the NIR flux peaks at a factor of 20 to 50 compared to quiescent emission \citep{GravityCollaboration2021}. For lower \bh spin values, since the stretching and opening of the separatrix is slower, the amount of time spent in the loading phase would be longer. For instance, for $a=0.6$, we could expect flux ropes to stay near the Y-ring twice longer \ie for 30-40$r_g/c$. Yet, it would still fall short of reaching the duration of the hot spots detected by \cite{Abuter2018b}. In addition, the Y-ring would be further and would thus probe regions of lower magnetic field, which would decrease the synchrotron power. The dynamics of the accretion disk itself seems to yield longer flux rope lifetimes in the GRMHD simulations of \cite{Nathanail2022} but it might be partly due to their artificially low reconnection rate of 1\% induced by the fluid formalism. In conclusion, the quick decay of the synchrotron power we measure in our simulations is thus incompatible with one flux rope being the source of the flare. 

\subsubsection{Collection of adjacent flux ropes}

In addition to the timing argument, the possibility that a single flux rope would be responsible for a hot spot is challenged by the multiple flux ropes expected to form along the Y-ring. Indeed, in our idealized simulation, the reconnecting layer is fed by magnetic flux from an underlying distribution on the disk which is assumed to be axisymmetric: the closed magnetic loops coupling the \bh to the disk cover the full azimuthal extent. Within the cone-shaped current layer formed, flux ropes break up and the associated flux ropes radiate but the stochastic nature of this process and the number $N>30$ of macroscopic structures derived in Section\,\ref{sec:growth} lead to an emission which originates from structures distributed uniformly in azimuth. Seen face-on, we should not see a coherent motion of the centroid since the synchrotron power of the largest flux ropes is comparable. 


However, accounting for a more realistic magnetic flux distribution in the disk would reconcile the observed and theoretical flare durations. Indeed, if the disk advects a magnetic loop of finite azimuthal extension $\Delta \phi$ large enough to form a tearing unstable current layer, a chain of flux ropes will form and flow away from a Y-arc of azimuthal extent $\Delta \phi$ in rotation with the separatrix. As an illustration, we take $\Delta \phi = \pi/8$, a value coherent with the size of the erupting and highly magnetized bubbles observed in GRMHD simulations of the MAD state \citep{Dexter2020,Ripperda2022} and suspected to be triggered by the magnetic Rayleigh-Taylor (\aka interchange) instability \citep{Begelman2022,Chatterjee2022}. After a few light crossing times, once the flux ropes have merged into a few macroscopic structures, the typical number of flux ropes at the Y-arc is $N\sim 30 \Delta \phi /(2\pi)\sim 2$ and the cumulated peak synchrotron power, for 2 flux ropes forming simultaneously, is thus of a few $10^{35}$erg$\cdot$s$^{-1}$. As these 2 flux ropes flow away from the Y-arc, their emissivity drops but is quickly exceeded by the one of newly formed flux ropes at the Y-arc. In this model, the hot spot traces the Y-arc itself, its rotation with the separatrix and the formation of successive flux ropes. The synchrotron flare is sustained as long as the coupling magnetic loop remains coherent and feeds the reconnection. If this loop is linked to the restructuration of the magnetosphere through the relaxation of the saturated magnetic flux in the MAD state, this coherence time is of $100r_g/c$ (according to \ileyk{2.5D} GRPIC simulations) to $300r_g/c$ \citep[according to GRMHD simulations][]{Galishnikova2022}, which overlaps with the observed duration of a flare in \sgr. Over this lapse of time, the cumulated synchrotron emission is jointly produced by a few tens of flux ropes. An observational hint in favor of this scenario is the presence of $\sim$10 minutes \ileyk{(\ie 30$r_g/c$)} long sub-flares within the broader envelop of the flare \citep{Genzel2003,Dodds-Eden2011}. \ileyk{In this picture, each of these sub-flare would be produced by one flux rope and the net cumulated emission from successive flux ropes would yield the whole flare.} The stochastic properties of this higher frequency signal would then carry the information on the number of flux ropes involved in the process and how quickly their emissivity decays.

\section{Summary and perspectives}
\label{sec:summary}

We performed global 3D GRPIC simulations of the magnetosphere around a Kerr \bh. Our observational motivation stems from the NIR flares and moving hot spots around \sgr that we try to reproduce with synchrotron emission from particles accelerated through magnetic reconnection. In our model, the magnetic flux is brought and sustained by an aligned, axisymmetric, steady and perfectly conducting disk in prograde Keplerian rotation. The magnetosphere relaxes to a topology comparable to the one we had previously characterized in 2.5D fully axisymmetric simulations (\pp): open magnetic field lines threading either the event horizon (jet region) or the disk (wind region), and magnetic field lines anchored in the innermost regions of the disk but closing within the event horizon or the ergosphere. At the intersection of these three regions, a Y-shaped magnetic field topology, or Y-ring, forms. Beyond, the jet and wind regions are separated by a cone-shaped current layer where vivid magnetic reconnection takes place and accelerates electrons and positrons up to \ileyk{relativistic} Lorentz factors. The location of the Y-ring and the opening angle of the current layer are similar to what we measured in 2.5D and we expect this good match to hold for lower \bh spin values than the one we considered ($a=0.99$).

However, there are noticeable differences between 2.5D and 3D simulations. First, plasmoids azimuthally break up when we let the 2.5D simulations relax to a fully 3D equilibrium. They are replaced by 3D flux ropes which are loaded with relativistic plasma and elongated in the azimuthal $\boldsymbol{\hat{\phi}}$ and quasi-radial $\boldsymbol{\hat{l}}$ directions. Flux ropes are seeded at the Y-ring where they co-rotate with the separatrix and episodically detach as magnetic field lines inflate (due to the growing toroidal component) and reconnect (due to the tearing instability). The sudden release of the magnetic tension combined with the fast rotation drives the flux ropes outward at mildly relativistic bulk speed along a coiled trajectory around the jet. The current layer is more turbulent than in 2.5D, with supplementary drift-kink and tearing modes triggered in the azimuthal direction. In addition to the reconnection of the poloidal component of the magnetic field across the current layer, we now capture the reconnection of the azimuthal component whose contribution is comparable. In spite of the complementary amount of magnetic flux available for reconnection, we measure a reconnection rate $\beta_{\rm rec}\sim 7$\% similar to 2.5D simulations, probably because of the more turbulent and thicker current layer. Owing to the possible role played by the outer light surface in shaping the magnetosphere, we foresee that these results would remain essentially the same for more realistic radial profiles of magnetic flux distribution on the disk.

We analyzed the motion and synchrotron emission of the largest flux ropes in order to determine whether this model could reproduce the observations of moving hot spots during NIR flares from \sgr \citep{Abuter2018b}. As they form at the Y-ring and flow along the current layer once they detach, the flux ropes describe a spiral of increasing pitch angle when viewed face-on. The trajectory resembles the apparent orbit of the hot spots observed. Its scale and timing properties essentially depend on the \bh spin and they match the observations for a positive spin of 0.65 to 0.8. Interestingly, we notice that this range of \bh spin is in agreement with what \cite{Collaboration2022} deduced independently from the image of \sgr in quiescent state (0.5 to 0.94). The plasma-loaded flux ropes almost co-rotate with the stretched separatrix when it is still attached to the Y-ring. Consequently, the apparent super-Keplerian motion is due to the shift between the separatrix footpoint on the disk, which sets the angular speed, and the projected distance of the Y-ring to the spin axis. Most importantly, \ileyk{the magnetization $\sigma\sim 1,000$ we work with enables us to accelerate particles up to} Lorentz factors of 100 to 1,000, consistent with a synchrotron emission peaking in NIR for \sgr parameters. Also, the synchrotron power of a single flux rope peaks at 10$^{35}$erg$\cdot$s$^{-1}$ near the Y-ring, in agreement with the observed NIR flares. X-ray flares would be within reach for a higher magnetization and thus a higher maximum particle Lorentz factor.

However, the synchrotron power emitted by a single flux rope quickly fades away once it detaches from the Y-ring. Regardless what the \bh spin is, we are not able to sustain the observed NIR luminosity over 30 to 60 minutes (\ie 90 to 180 $r_g/c$). Furthermore, the open trajectories we obtain are in tension with the full loop in the (Q,U) polarization map observed by \cite{Wielgus2022}. We argue that an important missing ingredient of our model susceptible to solve these discrepancies relates to our idealized representation of the disk. Indeed, our 3D model relies on an axisymmetric magnetic structure in the underlying disk which acts as a passive and steady reservoir of flux over $\sim 100r_g/c$. As a consequence, flux ropes form all along the Y-ring, an additional drawback to explain the centroid shift captured by the GRAVITY collaboration. If instead, we account for the finite azimuthal extent of a coherent magnetic structure in the disk, then only a fraction of the Y-ring undergoes the mechanism we presented here. In this alternative scenario, the kinematic results remain the same (and so is our prediction for the \bh spin) but a hot spot would actually be the outcome of the cumulated emission from a few flux ropes successively forming on a segment of the Y-ring. The narrower the azimuthal extent of the underlying magnetic structure, the lower the instantaneous number of flux ropes involved in the process and the noisier the net flare's light curve. The duration of the flare would then be set by the coherence time scale in the disk. If the magnetic structure originates from an eruptive expulsion of magnetic flux as seen in GRMHD simulations of the MAD state \citep{Porth2021,Zhdankin2023}, this time scale would be of the order of 30 to 60 minutes in \sgr and the number of flux ropes' forming sites along the Y-ring would be of 3 to 5. The presence of sub-flares with a coherence time scale of $\sim$10 minutes in the flares of \sgr corroborates this model \citep{Genzel2003,Dodds-Eden2011} \ileyk{since it matches the coherence time scale of a single flux rope. Each sub-flare would be produced by one flux rope and the cumulated emission would yield the whole flare. An a posteriori radiative transfer computation with a ray-tracing code remains necessary to extract from these GRPIC simulations synthetic light curves, astrometric motion, (Q,U) polarization maps and spectra for direct comparison with observations \citep{Ball2021,Aimar2023}.}

In this mechanism, particles are preferentially accelerated at the Y-ring above the disk, near the basis of the electromagnetic jet. Acceleration sites in the innermost equatorial region of the accretion flow have also been identified in GRPIC \citep{Crinquand2021} and GRMHD simulations \citep{Ripperda2020}, although the dependence of their kinematic and radiative properties on the \bh spin is unclear. If they turn out to be preferentially found within the \isco, where the magnetic field is higher, transient emission sites could only reproduce the hot spots' distance to the \bh spin axis for lower spin values (\ie <0.6) since the \isco is narrower than the Y-ring. We showed in \pp that lower spins lead to much lower amounts of particle energies. Although the properties of the oblique and equatorial current layer differ, it remains to be investigated whether flux ropes in an equatorial current layer could reproduce simultaneously the kinematic and radiative properties of \sgr's hot spots for lower spin values. 







\begin{acknowledgements}
The authors wish to thank Kyle Parfrey for fruitful discussions on the dynamics of the magnetosphere. This project has received funding from the European Research Council (ERC) under the European Union’s Horizon 2020 research and innovation programme (grant agreement No 863412). Computing resources were provided by TGCC and CINES under the allocation A0130407669 made by GENCI. This research was facilitated by the Multimessenger Plasma Physics Center (MPPC), NSF grant PHY-2206607.
\end{acknowledgements}


\bibliographystyle{aa} 
\begin{tiny}
\bibliography{all_my_library}

\begin{thebibliography}{85}
\expandafter\ifx\csname natexlab\endcsname\relax\def\natexlab#1{#1}\fi

\bibitem[{Abuter {et~al.}(2020)Abuter, Amorim, Baub{\"{o}}ck, Berger, Bonnet,
  Brandner, Cardoso, Cl{\'{e}}net, {De Zeeuw}, Dexter, Eckart, Eisenhauer,
  {F{\"{o}}rster Schreiber}, Garcia, Gao, Gendron, Genzel, Gillessen, Habibi,
  Haubois, Henning, Hippler, Horrobin, Jim{\'{e}}nez-Rosales, Jochum, Jocou,
  Kaufer, Kervella, Lacour, Lapeyr{\`{e}}re, {Le Bouquin}, L{\'{e}}na, Nowak,
  Ott, Paumard, Perraut, Perrin, Pfuhl, Rodr{\'{i}}guez-Coira, Shangguan,
  Scheithauer, Stadler, Straub, Straubmeier, Sturm, Tacconi, Vincent, {Von
  Fellenberg}, Waisberg, Widmann, Wieprecht, Wiezorrek, Woillez, Yazici, \&
  Zins}]{Abuter2020}
Abuter, R., Amorim, A., Baub{\"{o}}ck, M., {et~al.} 2020, Astron. Astrophys.,
  636, L5

\bibitem[{Abuter {et~al.}(2021)Abuter, Amorim, Baub{\"{o}}ck, Berger, Bonnet,
  Brandner, Cl{\'{e}}net, Dallilar, Davies, {De Zeeuw}, Dexter, Drescher,
  Eisenhauer, {F{\"{o}}rster Schreiber}, Garcia, Gao, Gendron, Genzel,
  Gillessen, Habibi, Haubois, Hei{\ss}el, Henning, Hippler, Horrobin,
  Jim{\'{e}}nez-Rosales, Jochum, Jocou, Kaufer, Kervella, Lacour,
  Lapeyr{\`{e}}re, {Le Bouquin}, L{\'{e}}na, Lutz, Nowak, Ott, Paumard,
  Perraut, Perrin, Pfuhl, Rabien, Rodr{\'{i}}guez-Coira, Shangguan, Shimizu,
  Scheithauer, Stadler, Straub, Straubmeier, Sturm, Tacconi, Vincent, {Von
  Fellenberg}, Waisberg, Widmann, Wieprecht, Wiezorrek, Woillez, Yazici, \&
  Zins}]{Abuter2021}
Abuter, R., Amorim, A., Baub{\"{o}}ck, M., {et~al.} 2021, Astron. Astrophys.,
  645, A127

\bibitem[{Aimar {et~al.}(2023)Aimar, Dmytriiev, Vincent, Paumard, Perrin, \&
  Zech}]{Aimar2023}
Aimar, N., Dmytriiev, A., Vincent, F.~H., {et~al.} 2023, Astron. Astrophys.,
  arXiv:2301.11874

\bibitem[{Baganoff {et~al.}(2001)Baganoff, Bautz, Brandt, Chartas, Feigelson,
  Garmire, Maeda, Morris, Ricker, Townsley, \& Walter}]{Baganoff2001}
Baganoff, F.~K., Bautz, M.~W., Brandt, W.~N., {et~al.} 2001, Nature, 413, 45

\bibitem[{Ball {et~al.}(2021)Ball, {\"{O}}zel, Christian, Chan, \&
  Psaltis}]{Ball2021}
Ball, D., {\"{O}}zel, F., Christian, P., Chan, C.-K., \& Psaltis, D. 2021,
  Astrophys. J., 917, 8

\bibitem[{Ball {et~al.}(2018)Ball, Sironi, \& {\"{O}}zel}]{Ball2018}
Ball, D., Sironi, L., \& {\"{O}}zel, F. 2018, Astrophys. J., 862, 80

\bibitem[{Barkov \& Komissarov(2016)}]{Barkov2016}
Barkov, M.~V. \& Komissarov, S.~S. 2016, Mon. Not. R. Astron. Soc., 458, 1939

\bibitem[{Begelman {et~al.}(2022)Begelman, Scepi, \& Dexter}]{Begelman2022}
Begelman, M.~C., Scepi, N., \& Dexter, J. 2022, Mon. Not. R. Astron. Soc., 511,
  2040

\bibitem[{Blandford \& Znajek(1977)}]{Blandford1977}
Blandford, R.~D. \& Znajek, R.~L. 1977, Mon. Not. R. Astron. Soc., 179, 433

\bibitem[{Bower {et~al.}(2018)Bower, Broderick, Dexter, Doeleman, Falcke, Fish,
  Johnson, Marrone, Moran, Moscibrodzka, Peck, Plambeck, \& Rao}]{Bower2018}
Bower, G.~C., Broderick, A., Dexter, J., {et~al.} 2018, Astrophys. J., 868, 101

\bibitem[{Bower {et~al.}(2019)Bower, Dexter, Asada, Brinkerink, Falcke, Ho,
  Inoue, Markoff, Marrone, Matsushita, Moscibrodzka, Nakamura, Peck, \&
  Rao}]{Bower2019}
Bower, G.~C., Dexter, J., Asada, K., {et~al.} 2019, Astrophys. J., 881, L2

\bibitem[{Boyce {et~al.}(2019)Boyce, Haggard, Witzel, Willner, Neilsen, Hora,
  Markoff, Ponti, Baganoff, Becklin, Fazio, Lowrance, Morris, \&
  Smith}]{Boyce2019}
Boyce, H., Haggard, D., Witzel, G., {et~al.} 2019, Astrophys. J., 871, 161

\bibitem[{Cassak {et~al.}(2017)Cassak, Liu, \& Shay}]{Cassak2017}
Cassak, P.~A., Liu, Y.~H., \& Shay, M.~A. 2017, {A review of the 0.1
  reconnection rate problem}

\bibitem[{Cerutti \& Giacinti(2021)}]{Cerutti2021}
Cerutti, B. \& Giacinti, G. 2021, Astron. Astrophys., 656, A91

\bibitem[{Cerutti {et~al.}(2015)Cerutti, Philippov, Parfrey, \&
  Spitkovsky}]{Cerutti2015}
Cerutti, B., Philippov, A., Parfrey, K., \& Spitkovsky, A. 2015, Mon. Not. R.
  Astron. Soc., 448, 606

\bibitem[{Cerutti \& Philippov(2017)}]{Cerutti2017a}
Cerutti, B. \& Philippov, A.~A. 2017, Astron. Astrophys., 607, A134

\bibitem[{Cerutti {et~al.}(2013)Cerutti, Werner, Uzdensky, \&
  Begelman}]{Cerutti2013}
Cerutti, B., Werner, G.~R., Uzdensky, D.~A., \& Begelman, M.~C. 2013,
  Astrophys. J., 770, 147

\bibitem[{Cerutti {et~al.}(2014)Cerutti, Werner, Uzdensky, \&
  Begelman}]{Cerutti2014}
Cerutti, B., Werner, G.~R., Uzdensky, D.~A., \& Begelman, M.~C. 2014,
  Astrophys. J., 782, 104

\bibitem[{Chael {et~al.}(2023)Chael, Issaoun, Pesce, Johnson, Ricarte, Fromm,
  \& Mizuno}]{Chael2023}
Chael, A., Issaoun, S., Pesce, D.~W., {et~al.} 2023, Astrophys. J., 945, 40

\bibitem[{Chashkina {et~al.}(2021)Chashkina, Bromberg, Levinson, Chashkina,
  Bromberg, \& Levinson}]{Chashkina2021a}
Chashkina, A., Bromberg, O., Levinson, A., {et~al.} 2021, 508, 1241

\bibitem[{Chatterjee \& Narayan(2022)}]{Chatterjee2022}
Chatterjee, K. \& Narayan, R. 2022, Astrophys. J., 941, 30

\bibitem[{Comisso \& Asenjo(2021)}]{Comisso2021}
Comisso, L. \& Asenjo, F.~A. 2021, Phys. Rev. D, 103, 023014

\bibitem[{Comisso \& Sironi(2019)}]{Comisso2019}
Comisso, L. \& Sironi, L. 2019, Astrophys. J., 886, 122

\bibitem[{Crinquand {et~al.}(2021)Crinquand, Cerutti, Dubus, Parfrey, \&
  Philippov}]{Crinquand2021}
Crinquand, B., Cerutti, B., Dubus, G., Parfrey, K., \& Philippov, A. 2021,
  Astron. Astrophys., 650, A163

\bibitem[{Crinquand {et~al.}(2020)Crinquand, Cerutti, Philippov, Parfrey, \&
  Dubus}]{Crinquand2020}
Crinquand, B., Cerutti, B., Philippov, A., Parfrey, K., \& Dubus, G. 2020,
  Phys. Rev. Lett., 124 [\eprint[arXiv]{2003.03548}]

\bibitem[{Cruz-Osorio {et~al.}(2022)Cruz-Osorio, Fromm, Mizuno, Nathanail,
  Younsi, Porth, Davelaar, Falcke, Kramer, \& Rezzolla}]{Cruz-Osorio2022}
Cruz-Osorio, A., Fromm, C.~M., Mizuno, Y., {et~al.} 2022, Nat. Astron., 6, 103

\bibitem[{Dallilar {et~al.}(2022)Dallilar, {Von Fellenberg}, Bauboeck, {De
  Zeeuw}, Drescher, Eisenhauer, Genzel, Gillessen, Habibi, Ott, Ponti, Stadler,
  Straub, Widmann, Witzel, \& Young}]{Dallilar2022}
Dallilar, Y., {Von Fellenberg}, S., Bauboeck, M., {et~al.} 2022, Astron.
  Astrophys., 658, A111

\bibitem[{{De Gouveia Dal Pino} \& Lazarian(2005)}]{DeGouveiaDalPino2005}
{De Gouveia Dal Pino}, E.~M. \& Lazarian, A. 2005, Astron. Astrophys., 441, 845

\bibitem[{Dexter {et~al.}(2020{\natexlab{a}})Dexter, Jim{\'{e}}nez-Rosales,
  Ressler, Tchekhovskoy, Baub{\"{o}}ck, {De Zeeuw}, Eisenhauer, {Von
  Fellenberg}, Gao, Genzel, Gillessen, Habibi, Ott, Stadler, Straub, \&
  Widmann}]{Dexter2020a}
Dexter, J., Jim{\'{e}}nez-Rosales, A., Ressler, S.~M., {et~al.}
  2020{\natexlab{a}}, Mon. Not. R. Astron. Soc., 494, 4168

\bibitem[{Dexter {et~al.}(2020{\natexlab{b}})Dexter, Tchekhovskoy,
  Jim{\'{e}}nez-Rosales, Ressler, Baub{\"{o}}ck, Dallilar, {De Zeeuw},
  Eisenhauer, {Von Fellenberg}, Gao, Genzel, Gillessen, Habibi, Ott, Stadler,
  Straub, \& Widmann}]{Dexter2020}
Dexter, J., Tchekhovskoy, A., Jim{\'{e}}nez-Rosales, A., {et~al.}
  2020{\natexlab{b}}, Mon. Not. R. Astron. Soc., 497, 4999

\bibitem[{Dodds-Eden {et~al.}(2011)Dodds-Eden, Gillessen, Fritz, Eisenhauer,
  Trippe, Genzel, Ott, Bartko, Pfuhl, Bower, Goldwurm, Porquet, Trap, \&
  Yusef-Zadeh}]{Dodds-Eden2011}
Dodds-Eden, K., Gillessen, S., Fritz, T.~K., {et~al.} 2011, Astrophys. J., 728,
  37

\bibitem[{{El Mellah} {et~al.}(2022){El Mellah}, Cerutti, Crinquand, \&
  Parfrey}]{ElMellah2022}
{El Mellah}, I., Cerutti, B., Crinquand, B., \& Parfrey, K. 2022, Astron.
  Astrophys., 663, A169

\bibitem[{{Event Horizon Telescope Collaboration}
  {et~al.}(2022{\natexlab{a}}){Event Horizon Telescope Collaboration}, Akiyama,
  Alberdi, Alef, Algaba, Anantua, Asada, Azulay, Bach, Baczko, Ball,
  Balokovi{\'{c}}, Barrett, Baub{\"{o}}ck, Benson, Bintley, Blackburn,
  Blundell, Bouman, Bower, Boyce, Bremer, Brinkerink, Brissenden, Britzen,
  Broderick, Broguiere, Bronzwaer, Bustamante, Byun, Carlstrom, Ceccobello,
  Chael, Chan, Chatterjee, Chatterjee, Chen, Chen, Cheng, Cho, Christian,
  Conroy, Conway, Cordes, Crawford, Crew, Cruz-Osorio, Cui, Davelaar, {De
  Laurentis}, Deane, Dempsey, Desvignes, Dexter, Dhruv, Doeleman, Dougal, Dzib,
  Eatough, Emami, Falcke, Farah, Fish, Fomalont, Ford, Fraga-Encinas, Freeman,
  Friberg, Fromm, Fuentes, Galison, Gammie, Garc{\'{i}}a, Gentaz, Georgiev,
  Goddi, Gold, G{\'{o}}mez-Ruiz, G{\'{o}}mez, Gu, Gurwell, Hada, Haggard,
  Haworth, Hecht, Hesper, Heumann, Ho, Ho, Honma, Huang, Huang, Hughes, Ikeda,
  Impellizzeri, Inoue, Issaoun, James, Jannuzi, Janssen, Jeter, Jiang,
  Jim{\'{e}}nez-Rosales, Johnson, Jorstad, Joshi, Jung, Karami, Karuppusamy,
  Kawashima, Keating, Kettenis, Kim, Kim, Kim, Kim, Kino, Koay, Kocherlakota,
  Kofuji, Koch, Koyama, Kramer, Kramer, Krichbaum, Kuo, {La Bella}, Lauer, Lee,
  Lee, Leung, Levis, Li, Lico, Lindahl, Lindqvist, Lisakov, Liu, Liu, Liuzzo,
  Lo, Lobanov, Loinard, Lonsdale, Lu, Mao, Marchili, Markoff, Marrone,
  Marscher, Mart{\'{i}}-Vidal, Matsushita, Matthews, Medeiros, Menten,
  Michalik, Mizuno, Mizuno, Moran, Moriyama, Moscibrodzka, M{\"{u}}ller, Mus,
  Musoke, Myserlis, Nadolski, Nagai, Nagar, Nakamura, Narayan, Narayanan,
  Natarajan, Nathanail, Fuentes, Neilsen, Neri, Ni, Noutsos, Nowak, Oh, Okino,
  Olivares, Ortiz-Le{\'{o}}n, Oyama, {\"{O}}zel, Palumbo, Paraschos, Park,
  Parsons, Patel, Pen, Pesce, Pi{\'{e}}tu, Plambeck, PopStefanija, Porth,
  P{\"{o}}tzl, Prather, Preciado-L{\'{o}}pez, Psaltis, Pu, Ramakrishnan, Rao,
  Rawlings, Raymond, Rezzolla, Ricarte, Ripperda, Roelofs, Rogers, Ros,
  Romero-Ca{\~{n}}izales, Roshanineshat, Rottmann, Roy, Ruiz, Ruszczyk, Rygl,
  S{\'{a}}nchez, S{\'{a}}nchez-Arg{\"{u}}elles, S{\'{a}}nchez-Portal, Sasada,
  Satapathy, Savolainen, Schloerb, Schonfeld, Schuster, Shao, Shen, Small,
  Sohn, SooHoo, Souccar, Sun, Tazaki, Tetarenko, Tiede, Tilanus, Titus, Torne,
  Traianou, Trent, Trippe, Turk, van Bemmel, van Langevelde, van Rossum, Vos,
  Wagner, Ward-Thompson, Wardle, Weintroub, Wex, Wharton, Wielgus, Wiik,
  Witzel, Wondrak, Wong, Wu, Yamaguchi, Yoon, Young, Young, Younsi, Yuan, Yuan,
  Zensus, Zhang, Zhao, Zhao, Agurto, Allardi, Amestica, Araneda, Arriagada,
  Berghuis, Bertarini, Berthold, Blanchard, Brown, C{\'{a}}rdenas, Cantzler,
  Caro, Castillo-Dom{\'{i}}nguez, Chan, Chang, Chang, Chang, Chang, Chen,
  Chilson, Chuter, Ciechanowicz, Colin-Beltran, Coulson, Crowley, Degenaar,
  Dornbusch, Dur{\'{a}}n, Everett, Faber, Forster, Fuchs, Gale, Geertsema,
  Gonz{\'{a}}lez, Graham, Gueth, Halverson, Han, Han, Hasegawa,
  Hern{\'{a}}ndez-Rebollar, Herrera, Herrero-Illana, Heyminck, Hirota, Hoge,
  {Hostler Schimpf}, Howie, Huang, Jiang, Jinchi, John, Kimura, Klein, Kubo,
  Kuroda, Kwon, Lacasse, Laing, Leitch, Li, Liu, Liu, Lin, Lu, Mac-Auliffe,
  Martin-Cocher, Matulonis, Maute, Messias, Meyer-Zhao, Monta{\~{n}}a,
  Montenegro-Montes, Montgomerie, {Moreno Nolasco}, Muders, Nishioka, Norton,
  Nystrom, Ogawa, Olivares, Oshiro, P{\'{e}}rez-Beaupuits, Parra, Phillips,
  Poirier, Pradel, Qiu, Raffin, Rahlin, Ram{\'{i}}rez, Ressler, Reynolds,
  Rodr{\'{i}}guez-Montoya, Saez-Madain, Santana, Shaw, Shirkey, Silva, Snow,
  Sousa, Sridharan, Stahm, Stark, Test, Torstensson, Venegas, Walther, Wei,
  White, Wieching, Wijnands, Wouterloot, Yu, {Yu (于威)}, \&
  Zeballos}]{Collaboration2022}
{Event Horizon Telescope Collaboration}, Akiyama, K., Alberdi, A., {et~al.}
  2022{\natexlab{a}}, Astrophys. J. Lett. Vol. 930, Issue 2, id.L12,
  {\textless}NUMPAGES{\textgreater}21{\textless}/NUMPAGES{\textgreater} pp.,
  930, L12

\bibitem[{{Event Horizon Telescope Collaboration}
  {et~al.}(2022{\natexlab{b}}){Event Horizon Telescope Collaboration}, Akiyama,
  Alberdi, Alef, {Carlos Algaba}, Anantua, Asada, Azulay, Bach, Baczko, Ball,
  Balokovi{\'{c}}, Barrett, Baub{\"{o}}ck, Benson, Bintley, Blackburn,
  Blundell, Bouman, Bower, Boyce, Bremer, Brinkerink, Brissenden, Britzen,
  Broderick, Broguiere, Bronzwaer, Bustamante, Byun, Carlstrom, Ceccobello,
  Chael, Chan, Chatterjee, Chatterjee, Chen, Chen, Cheng, Cho, Christian,
  Conroy, Conway, Cordes, Crawford, Crew, Cruz-Osorio, Cui, Davelaar, {De
  Laurentis}, Deane, Dempsey, Desvignes, Dexter, Dhruv, Doeleman, Dougal, Dzib,
  Eatough, Emami, Falcke, Farah, Fish, Fomalont, Ford, Fraga-Encinas, Freeman,
  Friberg, Fromm, Fuentes, Galison, Gammie, Garc{\'{i}}a, Gentaz, Georgiev,
  Goddi, Gold, G{\'{o}}mez-Ruiz, G{\'{o}}mez, Gu, Gurwell, Hada, Haggard,
  Haworth, Hecht, Hesper, Heumann, Ho, Ho, Honma, Huang, Huang, Hughes, Ikeda,
  {Violette Impellizzeri}, Inoue, Issaoun, James, Jannuzi, Janssen, Jeter,
  Jiang, Jim{\'{e}}nez-Rosales, Johnson, Jorstad, Joshi, Jung, Karami,
  Karuppusamy, Kawashima, Keating, Kettenis, Kim, Kim, Kim, Kim, Kino, Koay,
  Kocherlakota, Kofuji, Koch, Koyama, Kramer, Kramer, Krichbaum, Kuo, Bella,
  Lauer, Lee, Lee, Leung, Levis, Li, Lico, Lindahl, Lindqvist, Lisakov, Liu,
  Liu, Liuzzo, Lo, Lobanov, Loinard, Lonsdale, Lu, Mao, Marchili, Markoff,
  Marrone, Marscher, Mart{\'{i}}-Vidal, Matsushita, Matthews, Medeiros, Menten,
  Michalik, Mizuno, Mizuno, Moran, Moriyama, Moscibrodzka, M{\"{u}}ller, Mus,
  Musoke, Myserlis, Nadolski, Nagai, Nagar, Nakamura, Narayan, Narayanan,
  Natarajan, Nathanail, {Navarro Fuentes}, Neilsen, Neri, Ni, Noutsos, Nowak,
  Oh, Okino, Olivares, Ortiz-Le{\'{o}}n, Oyama, {\"{O}}zel, Palumbo, {Filippos
  Paraschos}, Park, Parsons, Patel, Pen, Pesce, Pi{\'{e}}tu, Plambeck,
  PopStefanija, Porth, P{\"{o}}tzl, Prather, Preciado-L{\'{o}}pez, Psaltis, Pu,
  Ramakrishnan, Rao, Rawlings, Raymond, Rezzolla, Ricarte, Ripperda, Roelofs,
  Rogers, Ros, Romero-Ca{\~{n}}izales, Roshanineshat, Rottmann, Roy, Ruiz,
  Ruszczyk, Rygl, S{\'{a}}nchez, S{\'{a}}nchez-Arg{\"{u}}elles,
  S{\'{a}}nchez-Portal, Sasada, Satapathy, Savolainen, Schloerb, Schonfeld,
  Schuster, Shao, Shen, Small, Sohn, SooHoo, Souccar, Sun, Tazaki, Tetarenko,
  Tiede, Tilanus, Titus, Torne, Traianou, Trent, Trippe, Turk, van Bemmel, van
  Langevelde, van Rossum, Vos, Wagner, Ward-Thompson, Wardle, Weintroub, Wex,
  Wharton, Wielgus, Wiik, Witzel, Wondrak, Wong, Wu, Yamaguchi, Yoon, Young,
  Young, Younsi, Yuan, Yuan, Zensus, Zhang, Zhao, Zhao, Chan, Qiu, Ressler,
  White, Algaba, Anantua, Asada, Azulay, Bach, Baczko, Ball, Balokovi{\'{c}},
  Barrett, Baub{\"{o}}ck, Benson, Bintley, Blackburn, Blundell, Bouman, Bower,
  Boyce, Bremer, Brinkerink, Brissenden, Britzen, Broderick, Broguiere,
  Bronzwaer, Bustamante, Byun, Carlstrom, Ceccobello, Chael, Chan, Chatterjee,
  Chatterjee, Chen, Chen, Cheng, Cho, Christian, Conroy, Conway, Cordes,
  Crawford, Crew, Cruz-Osorio, Cui, Davelaar, {De Laurentis}, Deane, Dempsey,
  Desvignes, Dexter, Dhruv, Doeleman, Dougal, Dzib, Eatough, Emami, Falcke,
  Farah, Fish, Fomalont, Ford, Fraga-Encinas, Freeman, Friberg, Fromm, Fuentes,
  Galison, Gammie, Garc{\'{i}}a, Gentaz, Georgiev, Goddi, Gold,
  G{\'{o}}mez-Ruiz, G{\'{o}}mez, Gu, Gurwell, Hada, Haggard, Haworth, Hecht,
  Hesper, Heumann, Ho, Ho, Honma, Huang, Huang, Hughes, Ikeda, {Violette
  Impellizzeri}, Inoue, Issaoun, James, Jannuzi, Janssen, Jeter, Jiang,
  Jim{\'{e}}nez-Rosales, Johnson, Jorstad, Joshi, Jung, Karami, Karuppusamy,
  Kawashima, Keating, Kettenis, Kim, Kim, Kim, Kim, Kino, Koay, Kocherlakota,
  Kofuji, Koch, Koyama, Kramer, Kramer, Krichbaum, Kuo, {La Bella}, Lauer, Lee,
  Lee, Leung, Levis, Li, Lico, Lindahl, Lindqvist, Lisakov, Liu, Liu, Liuzzo,
  Lo, Lobanov, Loinard, Lonsdale, Lu, Mao, Marchili, Markoff, Marrone,
  Marscher, Mart{\'{i}}-Vidal, Matsushita, Matthews, Medeiros, Menten,
  Michalik, Mizuno, Mizuno, Moran, Moriyama, Moscibrodzka, M{\"{u}}ller, Mus,
  Musoke, Myserlis, Nadolski, Nagai, Nagar, Nakamura, Narayan, Narayanan,
  Natarajan, Nathanail, {Navarro Fuentes}, Neilsen, Neri, Ni, Noutsos, Nowak,
  Oh, Okino, Olivares, Ortiz-Le{\'{o}}n, Oyama, {\"{O}}zel, Palumbo, {Filippos
  Paraschos}, Park, Parsons, Patel, Pen, Pesce, Pi{\'{e}}tu, Plambeck,
  PopStefanija, Porth, P{\"{o}}tzl, Prather, Preciado-L{\'{o}}pez, Psaltis, Pu,
  Ramakrishnan, Rao, Rawlings, Raymond, Rezzolla, Ricarte, Ripperda, Roelofs,
  Rogers, Ros, Romero-Ca{\~{n}}izales, Roshanineshat, Rottmann, Roy, Ruiz,
  Ruszczyk, Rygl, S{\'{a}}nchez, S{\'{a}}nchez-Arg{\"{u}}elles,
  S{\'{a}}nchez-Portal, Sasada, Satapathy, Savolainen, Schloerb, Schonfeld,
  Schuster, Shao, Shen, Small, Sohn, SooHoo, Souccar, Sun, Tazaki, Tetarenko,
  Tiede, Tilanus, Titus, Torne, Traianou, Trent, Trippe, Turk, van Bemmel, van
  Langevelde, van Rossum, Vos, Wagner, Ward-Thompson, Wardle, Weintroub, Wex,
  Wharton, Wielgus, Wiik, Witzel, Wondrak, Wong, Wu, Yamaguchi, Yoon, Young,
  Young, Younsi, Yuan, Yuan, Zensus, Zhang, Zhao, Zhao, Chan, Qiu, Ressler,
  White, Collaboration, Akiyama, Alberdi, Alef, Algaba, Anantua, Asada, Azulay,
  Bach, Baczko, Ball, Balokovi{\'{c}}, Barrett, Baub{\"{o}}ck, Benson, Bintley,
  Blackburn, Blundell, Bouman, Bower, Boyce, Bremer, Brinkerink, Brissenden,
  Britzen, Broderick, Broguiere, Bronzwaer, Bustamante, Byun, Carlstrom,
  Ceccobello, Chael, Chan, Chatterjee, Chatterjee, Chen, Chen, Cheng, Cho,
  Christian, Conroy, Conway, Cordes, Crawford, Crew, Cruz-Osorio, Cui,
  Davelaar, {De Laurentis}, Deane, Dempsey, Desvignes, Dexter, Dhruv, Doeleman,
  Dougal, Dzib, Eatough, Emami, Falcke, Farah, Fish, Fomalont, Ford,
  Fraga-Encinas, Freeman, Friberg, Fromm, Fuentes, Galison, Gammie,
  Garc{\'{i}}a, Gentaz, Georgiev, Goddi, Gold, G{\'{o}}mez-Ruiz, G{\'{o}}mez,
  Gu, Gurwell, Hada, Haggard, Haworth, Hecht, Hesper, Heumann, Ho, Ho, Honma,
  Huang, Huang, Hughes, Ikeda, {Violette Impellizzeri}, Inoue, Issaoun, James,
  Jannuzi, Janssen, Jeter, Jiang, Jim{\'{e}}nez-Rosales, Johnson, Jorstad,
  Joshi, Jung, Karami, Karuppusamy, Kawashima, Keating, Kettenis, Kim, Kim,
  Kim, Kim, Kino, Koay, Kocherlakota, Kofuji, Koch, Koyama, Kramer, Kramer,
  Krichbaum, Kuo, {La Bella}, Lauer, Lee, Lee, Leung, Levis, Li, Lico, Lindahl,
  Lindqvist, Lisakov, Liu, Liu, Liuzzo, Lo, Lobanov, Loinard, Lonsdale, Lu,
  Mao, Marchili, Markoff, Marrone, Marscher, Mart{\'{i}}-Vidal, Matsushita,
  Matthews, Medeiros, Menten, Michalik, Mizuno, Mizuno, Moran, Moriyama,
  Moscibrodzka, M{\"{u}}ller, Mus, Musoke, Myserlis, Nadolski, Nagai, Nagar,
  Nakamura, Narayan, Narayanan, Natarajan, Nathanail, {Navarro Fuentes},
  Neilsen, Neri, Ni, Noutsos, Nowak, Oh, Okino, Olivares, Ortiz-Le{\'{o}}n,
  Oyama, {\"{O}}zel, Palumbo, {Filippos Paraschos}, Park, Parsons, Patel, Pen,
  Pesce, Pi{\'{e}}tu, Plambeck, PopStefanija, Porth, P{\"{o}}tzl, Prather,
  Preciado-L{\'{o}}pez, Psaltis, Pu, Ramakrishnan, Rao, Rawlings, Raymond,
  Rezzolla, Ricarte, Ripperda, Roelofs, Rogers, Ros, Romero-Ca{\~{n}}izales,
  Roshanineshat, Rottmann, Roy, Ruiz, Ruszczyk, Rygl, S{\'{a}}nchez,
  S{\'{a}}nchez-Arg{\"{u}}elles, S{\'{a}}nchez-Portal, Sasada, Satapathy,
  Savolainen, Schloerb, Schonfeld, Schuster, Shao, Shen, Small, Sohn, SooHoo,
  Souccar, Sun, Tazaki, Tetarenko, Tiede, Tilanus, Titus, Torne, Traianou,
  Trent, Trippe, Turk, van Bemmel, van Langevelde, van Rossum, Vos, Wagner,
  Ward-Thompson, Wardle, Weintroub, Wex, Wharton, Wielgus, Wiik, Witzel,
  Wondrak, Wong, Wu, Yamaguchi, Yoon, Young, Young, Younsi, Yuan, Yuan, Zensus,
  Zhang, Zhao, Zhao, Chan, Qiu, Ressler, \&
  White}]{EventHorizonTelescopeCollaboration2022}
{Event Horizon Telescope Collaboration}, Akiyama, K., Alberdi, A., {et~al.}
  2022{\natexlab{b}}, Astrophys. J. Lett., 930, L16

\bibitem[{Galishnikova {et~al.}(2022)Galishnikova, Philippov, Quataert,
  Bacchini, Parfrey, \& Ripperda}]{Galishnikova2022}
Galishnikova, A., Philippov, A., Quataert, E., {et~al.} 2022, arXiv,
  arXiv:2212.02583

\bibitem[{Genzel {et~al.}(2003)Genzel, Sch{\"{o}}del, Ott, Eckart, Alexander,
  Lacombe, Rouan, \& Aschenbach}]{Genzel2003}
Genzel, R., Sch{\"{o}}del, R., Ott, T., {et~al.} 2003, Nature, 425, 934

\bibitem[{Ghez {et~al.}(2004)Ghez, Wright, Matthews, Thompson, {Le Mignant},
  Tanner, Hornstein, Morris, Becklin, \& Soifer}]{Ghez2004}
Ghez, A.~M., Wright, S.~A., Matthews, K., {et~al.} 2004, Astrophys. J., 601,
  L159

\bibitem[{Gillessen {et~al.}(2009)Gillessen, Eisenhauer, Fritz, Bartko,
  Dodds-Eden, Pfuhl, Ott, \& Genzel}]{Gillessen2009}
Gillessen, S., Eisenhauer, F., Fritz, T.~K., {et~al.} 2009, Astrophys. J., 707,
  L114

\bibitem[{Goldreich \& Julian(1969)}]{Goldreich1969}
Goldreich, P. \& Julian, W. 1969, Astrophys. J., 157, 9

\bibitem[{Goodbred \& Liu(2022)}]{Goodbred2023}
Goodbred, M. \& Liu, Y.~H. 2022, Phys. Rev. Lett., 129, 265101

\bibitem[{{Gravity Collaboration} {et~al.}(2018{\natexlab{a}}){Gravity
  Collaboration}, Abuter, Amorim, Anugu, Baub{\"{o}}ck, Benisty, Berger, Blind,
  Bonnet, Brandner, Buron, Collin, Chapron, Cl{\'{e}}net, {Du Foresto}, {De
  Zeeuw}, Deen, Delplancke-Str{\"{o}}bele, Dembet, Dexter, Duvert, Eckart,
  Eisenhauer, Finger, Schreiber, F{\'{e}}dou, Garcia, Lopez, Gao, Gendron,
  Genzel, Gillessen, Gordo, Habibi, Haubois, Haug, Hau{\ss}mann, Henning,
  Hippler, Horrobin, Hubert, Hubin, Rosales, Jochum, Jocou, Kaufer, Kellner,
  Kendrew, Kervella, Kok, Kulas, Lacour, Lapeyr{\`{e}}re, Lazareff, {Le
  Bouquin}, L{\'{e}}na, Lippa, Lenzen, M{\'{e}}rand, M{\"{u}}ler, Neumann, Ott,
  Palanca, Paumard, Pasquini, Perraut, Perrin, Pfuhl, Plewa, Rabien,
  Ram{\'{i}}rez, Ramos, Rau, Rodr{\'{i}}guez-Coira, Rohloff, Rousset,
  Sanchez-Bermudez, Scheithauer, Sch{\"{o}}ller, Schuler, Spyromilio, Straub,
  Straubmeier, Sturm, Tacconi, Tristram, Vincent, {Von Fellenberg}, Wank,
  Waisberg, Widmann, Wieprecht, Wiest, Wiezorrek, Woillez, Yazici, Ziegler, \&
  Zins}]{Abuter2018c}
{Gravity Collaboration}, Abuter, R., Amorim, A., {et~al.} 2018{\natexlab{a}},
  Astron. Astrophys., 615, L15

\bibitem[{{Gravity Collaboration} {et~al.}(2021){Gravity Collaboration},
  Abuter, Amorim, Baub{\"{o}}ck, Baganoff, Berger, Boyce, Bonnet, Brandner,
  Cl{\'{e}}net, Davies, de~Zeeuw, Dexter, Dallilar, Drescher, Eckart,
  Eisenhauer, Fazio, {F{\"{o}}rster Schreiber}, Foster, Gammie, Garcia, Gao,
  Gendron, Genzel, Ghisellini, Gillessen, Gurwell, Habibi, Haggard, Hailey,
  Harrison, Haubois, Hei{\ss}el, Henning, Hippler, Hora, Horrobin,
  Jim{\'{e}}nez-Rosales, Jochum, Jocou, Kaufer, Kervella, Lacour,
  Lapeyr{\`{e}}re, {Le Bouquin}, L{\'{e}}na, Lowrance, Lutz, Markoff, Mori,
  Morris, Neilsen, Nowak, Ott, Paumard, Perraut, Perrin, Ponti, Pfuhl, Rabien,
  Rodr{\'{i}}guez-Coira, Shangguan, Shimizu, Scheithauer, Smith, Stadler,
  Stern, Straub, Straubmeier, Sturm, Tacconi, Vincent, von Fellenberg,
  Waisberg, Widmann, Wieprecht, Wiezorrek, Willner, Witzel, Woillez, Yazici,
  Young, Zhang, Zins, Collaboration, Abuter, Amorim, Baub{\"{o}}ck, Baganoff,
  Berger, Boyce, Bonnet, Brandner, Cl{\'{e}}net, Davies, de~Zeeuw, Dexter,
  Dallilar, Drescher, Eckart, Eisenhauer, Fazio, {F{\"{o}}rster Schreiber},
  Foster, Gammie, Garcia, Gao, Gendron, Genzel, Ghisellini, Gillessen, Gurwell,
  Habibi, Haggard, Hailey, Harrison, Haubois, Hei{\ss}el, Henning, Hippler,
  Hora, Horrobin, Jim{\'{e}}nez-Rosales, Jochum, Jocou, Kaufer, Kervella,
  Lacour, Lapeyr{\`{e}}re, {Le Bouquin}, L{\'{e}}na, Lowrance, Lutz, Markoff,
  Mori, Morris, Neilsen, Nowak, Ott, Paumard, Perraut, Perrin, Ponti, Pfuhl,
  Rabien, Rodr{\'{i}}guez-Coira, Shangguan, Shimizu, Scheithauer, Smith,
  Stadler, Stern, Straub, Straubmeier, Sturm, Tacconi, Vincent, von Fellenberg,
  Waisberg, Widmann, Wieprecht, Wiezorrek, Willner, Witzel, Woillez, Yazici,
  Young, Zhang, \& Zins}]{GravityCollaboration2021}
{Gravity Collaboration}, Abuter, R., Amorim, A., {et~al.} 2021, A{\&}A, 654,
  A22

\bibitem[{{Gravity Collaboration} {et~al.}(2018{\natexlab{b}}){Gravity
  Collaboration}, Abuter, Amorim, Baub{\"{o}}ck, Berger, Bonnet, Brandner,
  Cl{\'{e}}net, {Du Foresto}, {De Zeeuw}, Deen, Dexter, Duvert, Eckart,
  Eisenhauer, Schreiber, Garcia, Gao, Gendron, Genzel, Gillessen, Guajardo,
  Habibi, Haubois, Henning, Hippler, Horrobin, Huber, Jim{\'{e}}nez-Rosales,
  Jocou, Kervella, Lacour, Lapeyr{\`{e}}re, Lazareff, {Le Bouquin}, L{\'{e}}na,
  Lippa, Ott, Panduro, Paumard, Perraut, Perrin, Pfuhl, Plewa, Rabien,
  Rodr{\'{i}}guez-Coira, Rousset, Sternberg, Straub, Straubmeier, Sturm,
  Tacconi, Vincent, {Von Fellenberg}, Waisberg, Widmann, Wieprecht, Wiezorrek,
  Woillez, Yazici, Anugu, Baub{\"{o}}ck, Benisty, Berger, Blind, Bonnet,
  Brandner, Buron, Collin, Chapron, Cl{\'{e}}net, {Du Foresto}, {De Zeeuw},
  Deen, Delplancke-Str{\"{o}}bele, Dembet, Dexter, Duvert, Eckart, Eisenhauer,
  Finger, Schreiber, F{\'{e}}dou, Garcia, Lopez, Gao, Gendron, Genzel,
  Gillessen, Gordo, Habibi, Haubois, Haug, Hau{\ss}mann, Henning, Hippler,
  Horrobin, Hubert, Hubin, Rosales, Jochum, Jocou, Kaufer, Kellner, Kendrew,
  Kervella, Kok, Kulas, Lacour, Lapeyr{\`{e}}re, Lazareff, {Le Bouquin},
  L{\'{e}}na, Lippa, Lenzen, M{\'{e}}rand, M{\"{u}}ler, Neumann, Ott, Palanca,
  Paumard, Pasquini, Perraut, Perrin, Pfuhl, Plewa, Rabien, Ram{\'{i}}rez,
  Ramos, Rau, Rodr{\'{i}}guez-Coira, Rohloff, Rousset, Sanchez-Bermudez,
  Scheithauer, Sch{\"{o}}ller, Schuler, Spyromilio, Straub, Straubmeier, Sturm,
  Tacconi, Tristram, Vincent, {Von Fellenberg}, Wank, Waisberg, Widmann,
  Wieprecht, Wiest, Wiezorrek, Woillez, Yazici, Ziegler, \& Zins}]{Abuter2018b}
{Gravity Collaboration}, Abuter, R., Amorim, A., {et~al.} 2018{\natexlab{b}},
  Astron. Astrophys., 618, 10

\bibitem[{{Gravity Collaboration} {et~al.}(2020){Gravity Collaboration},
  Baub{\"{o}}ck, Dexter, Abuter, Amorim, Berger, Bonnet, Brandner,
  Cl{\'{e}}net, {Coud{\'{e}} Du Foresto}, {De Zeeuw}, Duvert, Eckart,
  Eisenhauer, {F{\"{o}}rster Schreiber}, Gao, Garcia, Gendron, Genzel, Gerhard,
  Gillessen, Habibi, Haubois, Henning, Hippler, Horrobin,
  Jim{\'{e}}nez-Rosales, Jocou, Kervella, Lacour, Lapeyr{\`{e}}re, {Le
  Bouquin}, L{\'{e}}na, Ott, Paumard, Perraut, Perrin, Pfuhl, Rabien,
  {Rodriguez Coira}, Rousset, Scheithauer, Stadler, Sternberg, Straub,
  Straubmeier, Sturm, Tacconi, Vincent, {Von Fellenberg}, Waisberg, Widmann,
  Wieprecht, Wiezorrek, Woillez, \& Yazici}]{Baubock2020}
{Gravity Collaboration}, Baub{\"{o}}ck, M., Dexter, J., {et~al.} 2020, Astron.
  Astrophys., 635, 1

\bibitem[{Guo {et~al.}(2015)Guo, Liu, Daughton, \& Li}]{Guo2015}
Guo, F., Liu, Y.~H., Daughton, W., \& Li, H. 2015, Astrophys. J., 806
  [\eprint[arXiv]{1504.02193}]

\bibitem[{Jim{\'{e}}nez-Rosales {et~al.}(2020)Jim{\'{e}}nez-Rosales, Dexter,
  Widmann, Baub{\"{o}}ck, Abuter, Amorim, Berger, Bonnet, Brandner,
  Cl{\'{e}}net, {De Zeeuw}, Eckart, Eisenhauer, {F{\"{o}}rster Schreiber},
  Garcia, Gao, Gendron, Genzel, Gillessen, Habibi, Haubois, Hei{\ss}el,
  Henning, Hippler, Horrobin, Jochum, Jocou, Kaufer, Kervella, Lacour,
  Lapeyr{\`{e}}re, {Le Bouquin}, L{\'{e}}na, Nowak, Ott, Paumard, Perraut,
  Perrin, Pfuhl, Rodr{\'{i}}guez-Coira, Shangguan, Scheithauer, Stadler,
  Straub, Straubmeier, Sturm, Tacconi, Vincent, {Von Fellenberg}, Waisberg,
  Wieprecht, Wiezorrek, Woillez, Yazici, Zins, {Gravity Collaboration},
  Jim{\'{e}}nez-Rosales, Dexter, Widmann, Baub{\"{o}}ck, Abuter, Amorim,
  Berger, Bonnet, Brandner, Cl{\'{e}}net, {De Zeeuw}, Eckart, Eisenhauer,
  {F{\"{o}}rster Schreiber}, Garcia, Gao, Gendron, Genzel, Gillessen, Habibi,
  Haubois, Hei{\ss}el, Henning, Hippler, Horrobin, Jochum, Jocou, Kaufer,
  Kervella, Lacour, Lapeyr{\`{e}}re, {Le Bouquin}, L{\'{e}}na, Nowak, Ott,
  Paumard, Perraut, Perrin, Pfuhl, Rodr{\'{i}}guez-Coira, Shangguan,
  Scheithauer, Stadler, Straub, Straubmeier, Sturm, Tacconi, Vincent, {Von
  Fellenberg}, Waisberg, Wieprecht, Wiezorrek, Woillez, Yazici, \&
  Zins}]{Jimenez-Rosales2020}
Jim{\'{e}}nez-Rosales, A., Dexter, J., Widmann, F., {et~al.} 2020, Astron.
  Astrophys., 643, A56

\bibitem[{Kerr(1963)}]{Kerr1963}
Kerr, R.~P. 1963, Phys. Rev. Lett., 11, 237

\bibitem[{Komissarov(2004)}]{Komissarov2004a}
Komissarov, S.~S. 2004, Mon. Not. R. Astron. Soc., 350, 427

\bibitem[{Lin {et~al.}(2023)Lin, Li, \& Yuan}]{Lin2023}
Lin, X., Li, Y.-P., \& Yuan, F. 2023, Mon. Not. R. Astron. Soc., 520, 1271

\bibitem[{Loureiro {et~al.}(2012)Loureiro, Samtaney, Schekochihin, \&
  Uzdensky}]{Loureiro2012a}
Loureiro, N.~F., Samtaney, R., Schekochihin, A.~A., \& Uzdensky, D.~A. 2012,
  Phys. Plasmas, 19, 042303

\bibitem[{MacDonald \& Thorne(1982)}]{MacDonald1982}
MacDonald, D. \& Thorne, K.~S. 1982, Mon. Not. R. Astron. Soc., 198, 345

\bibitem[{Meringolo {et~al.}(2023)Meringolo, Cruz-Osorio, Rezzolla, \&
  Servidio}]{Meringolo2023}
Meringolo, C., Cruz-Osorio, A., Rezzolla, L., \& Servidio, S. 2023, Astrophys.
  J., 944, 122

\bibitem[{Nathanail {et~al.}(2020)Nathanail, Fromm, Porth, Olivares, Younsi,
  Mizuno, \& Rezzolla}]{Nathanail2020}
Nathanail, A., Fromm, C.~M., Porth, O., {et~al.} 2020, 495

\bibitem[{Nathanail {et~al.}(2022)Nathanail, Mpisketzis, Porth, Fromm, \&
  Rezzolla}]{Nathanail2022}
Nathanail, A., Mpisketzis, V., Porth, O., Fromm, C.~M., \& Rezzolla, L. 2022,
  Mon. Not. R. Astron. Soc., 513, 4267

\bibitem[{Parfrey {et~al.}(2019)Parfrey, Philippov, \& Cerutti}]{Parfrey2019}
Parfrey, K., Philippov, A., \& Cerutti, B. 2019, Phys. Rev. Lett., 122, 35101

\bibitem[{Philippov {et~al.}(2019)Philippov, Uzdensky, Spitkovsky, \&
  Cerutti}]{Philippov2019}
Philippov, A., Uzdensky, D.~A., Spitkovsky, A., \& Cerutti, B. 2019, Astrophys.
  J., 876, L6

\bibitem[{Ponti {et~al.}(2017)Ponti, George, Scaringi, Zhang, Jin, Dexter,
  Terrier, Clavel, Degenaar, Eisenhauer, Genzel, Gillessen, Goldwurm, Habibi,
  Haggard, Hailey, Harrison, Merloni, Mori, Nandra, Ott, Pfuhl, Plewa, \&
  Waisberg}]{Ponti2017}
Ponti, G., George, E., Scaringi, S., {et~al.} 2017, Mon. Not. R. Astron. Soc.,
  468, 2447

\bibitem[{Porth {et~al.}(2021)Porth, Mizuno, Younsi, \& Fromm}]{Porth2021}
Porth, O., Mizuno, Y., Younsi, Z., \& Fromm, C.~M. 2021, Mon. Not. R. Astron.
  Soc., 502, 2023

\bibitem[{Ressler {et~al.}(2020)Ressler, White, Quataert, \&
  Stone}]{Ressler2020a}
Ressler, S.~M., White, C.~J., Quataert, E., \& Stone, J.~M. 2020, Astrophys.
  J., 896, L6

\bibitem[{Ripperda {et~al.}(2020)Ripperda, Bacchini, \&
  Philippov}]{Ripperda2020}
Ripperda, B., Bacchini, F., \& Philippov, A.~A. 2020, Astrophys. J., 900, 100

\bibitem[{Ripperda {et~al.}(2022)Ripperda, Liska, Chatterjee, Musoke,
  Philippov, Markoff, Tchekhovskoy, \& Younsi}]{Ripperda2022}
Ripperda, B., Liska, M., Chatterjee, K., {et~al.} 2022, Astrophys. J. Lett.,
  924, L32

\bibitem[{Rowan {et~al.}(2017)Rowan, Sironi, \& Narayan}]{Rowan2017}
Rowan, M.~E., Sironi, L., \& Narayan, R. 2017, Astrophys. J., 850, 29

\bibitem[{Scepi {et~al.}(2022)Scepi, Dexter, \& Begelman}]{Scepi2022}
Scepi, N., Dexter, J., \& Begelman, M.~C. 2022, Mon. Not. R. Astron. Soc., 511,
  3536

\bibitem[{Sch{\"{o}}del {et~al.}(2009)Sch{\"{o}}del, Merritt, \&
  Eckart}]{Schoedel2009}
Sch{\"{o}}del, R., Merritt, D., \& Eckart, A. 2009, Astron. Astrophys., 502, 91

\bibitem[{Sironi {et~al.}(2016)Sironi, Giannios, \& Petropoulou}]{Sironi2016a}
Sironi, L., Giannios, D., \& Petropoulou, M. 2016, Mon. Not. R. Astron. Soc.,
  462, 48

\bibitem[{Sironi \& Spitkovsky(2014)}]{Sironi2014}
Sironi, L. \& Spitkovsky, A. 2014, Astrophys. J. Lett., 783, L21

\bibitem[{Tchekhovskoy {et~al.}(2011)Tchekhovskoy, Narayan, \&
  Mckinney}]{Tchekhovskoy2011}
Tchekhovskoy, A., Narayan, R., \& Mckinney, J.~C. 2011, Mon. Not. R. Astron.
  Soc. Lett., 418, L79

\bibitem[{Uzdensky(2005)}]{Uzdensky2005}
Uzdensky, D.~A. 2005, Astrophys. J., 620, 889

\bibitem[{Vos {et~al.}(2022)Vos, Moacibrodzka, \& Wielgus}]{Vos2022}
Vos, J., Moacibrodzka, M.~A., \& Wielgus, M. 2022, Astron. Astrophys., 668,
  A185

\bibitem[{Werner \& Uzdensky(2017)}]{Werner2017}
Werner, G.~R. \& Uzdensky, D.~A. 2017, Astrophys. J., 843, L27

\bibitem[{Werner \& Uzdensky(2021)}]{Werner2021}
Werner, G.~R. \& Uzdensky, D.~A. 2021, J. Plasma Phys., 87, 905870613

\bibitem[{Werner {et~al.}(2018)Werner, Uzdensky, Begelman, Cerutti, \&
  Nalewajko}]{Werner2018}
Werner, G.~R., Uzdensky, D.~A., Begelman, M.~C., Cerutti, B., \& Nalewajko, K.
  2018, Mon. Not. R. Astron. Soc., 473, 4840

\bibitem[{Werner {et~al.}(2015)Werner, Uzdensky, Cerutti, Nalewajko, \&
  Begelman}]{Werner2015}
Werner, G.~R., Uzdensky, D.~A., Cerutti, B., Nalewajko, K., \& Begelman, M.~C.
  2015, Astrophys. J., 816, L8

\bibitem[{Wielgus {et~al.}(2022)Wielgus, Moscibrodzka, Vos, Gelles,
  Mart{\'{i}}-Vidal, Farah, Marchili, Goddi, \& Messias}]{Wielgus2022}
Wielgus, M., Moscibrodzka, M., Vos, J., {et~al.} 2022, Astron. Astrophys., 665,
  arXiv:2209.09926

\bibitem[{Wong {et~al.}(2022)Wong, Prather, Dhruv, Ryan, Mo{\'{s}}cibrodzka,
  Chan, Joshi, Yarza, Ricarte, Shiokawa, Dolence, Noble, McKinney, \&
  Gammie}]{Wong2022}
Wong, G.~N., Prather, B.~S., Dhruv, V., {et~al.} 2022, Astrophys. J. Suppl.
  Ser., 259, 64

\bibitem[{Yuan {et~al.}(2009)Yuan, Lin, Wu, \& Ho}]{Yuan2009}
Yuan, F., Lin, J., Wu, K., \& Ho, L.~C. 2009, Mon. Not. R. Astron. Soc., 395,
  2183

\bibitem[{Yuan {et~al.}(2004)Yuan, Quataert, \& Narayan}]{Yuan2004}
Yuan, F., Quataert, E., \& Narayan, R. 2004, Astrophys. J., 606, 894

\bibitem[{Yuan {et~al.}(2019)Yuan, Blandford, \& Wilkins}]{Yuan2019b}
Yuan, Y., Blandford, R.~D., \& Wilkins, D.~R. 2019, Mon. Not. R. Astron. Soc.,
  484, 4920

\bibitem[{Zelenyi \& Krasnoselskikh(1979)}]{Zelenyi1979}
Zelenyi, L.~M. \& Krasnoselskikh, V.~V. 1979, Sov. Astron. Vol. 23, P. 460,
  1979, 23, 460

\bibitem[{Zenitani \& Hoshino(2007)}]{Zenitani2007}
Zenitani, S. \& Hoshino, M. 2007, Astrophys. J., 670, 702

\bibitem[{Zhang {et~al.}(2021)Zhang, Sironi, \& Giannios}]{Zhang2021}
Zhang, H., Sironi, L., \& Giannios, D. 2021, Astrophys. J., 922, 261

\bibitem[{Zhdankin {et~al.}(2023)Zhdankin, Ripperda, \&
  Philippov}]{Zhdankin2023}
Zhdankin, V., Ripperda, B., \& Philippov, A.~A. 2023, eprint arXiv:2302.05276,
  arXiv:2302.05276

\bibitem[{Zhou {et~al.}(2019)Zhou, Bhat, Loureiro, \& Uzdensky}]{Zhou2019}
Zhou, M., Bhat, P., Loureiro, N.~F., \& Uzdensky, D.~A. 2019, Phys. Rev. Res.,
  1, 012004

\bibitem[{Zhou {et~al.}(2020)Zhou, Loureiro, \& Uzdensky}]{Zhou2020}
Zhou, M., Loureiro, N.~F., \& Uzdensky, D.~A. 2020, J. Plasma Phys., 86,
  535860401

\bibitem[{Zhou {et~al.}(2021)Zhou, Wu, Loureiro, \& Uzdensky}]{Zhou2021}
Zhou, M., Wu, D.~H., Loureiro, N.~F., \& Uzdensky, D.~A. 2021, J. Plasma Phys.,
  87, 905870620

\end{thebibliography}
\end{tiny}

\begin{appendix}

\end{appendix}

\end{document}